\documentclass[12pt]{article}
\newcommand{\nc}{\newcommand}
\nc{\bib}{\bibitem}
\nc{\al}{\alpha}
\nc{\g}{\gamma}
\nc{\G}{\Gamma}
\nc{\D}{\Delta}
\nc{\eps}{\epsilon}
\nc{\la}{\lambda}
\nc{\La}{\Lambda}
\nc{\var}{\varphi}
\nc{\hn}{h^\vee}
\nc{\kn}{k^\vee}
\nc{\adg}{a^\dagger}
\nc{\bdg}{b^\dagger}
\nc{\ba}{\beta_\al}
\nc{\ga}{\g^\al}
\nc{\vpp}{{V_+}^+}
\nc{\cpp}{{C_+}^+}
\nc{\Vm}{V_{-\al^-}^{\al_1}}
\nc{\Vp}{V_{-\al^+}^{\al_1}}
\nc{\Vmb}{V_{-\beta^-}^{\al_1}}
\nc{\Gb}{\overline{G}}
\nc{\Gbc}{\overline{{\cal G}}}
\nc{\pa}{\partial}
\nc{\nn}{\nonumber \\ }
\nc{\hf}{\frac{1}{2}}         
\nc{\dz}{\frac{dz}{2\pi i}}
\nc{\fabc}{{f_{a,b}}^c}
\nc{\bin}[2]{\left (\begin{array}{c} {#1}\\ {#2} \end{array}\right )}
\nc{\ben}{\begin{equation}}
\nc{\een}{\end{equation}}
\nc{\bea}{\begin{eqnarray}}
\nc{\eea}{\end{eqnarray}}
\nc{\bra}[1]{\langle {#1}|}
\nc{\ket}[1]{|{#1}\rangle}
\newcommand{\Z}{\mbox{$Z\hspace{-2mm}Z$}}
\nc{\C}{\mbox{\hspace{1.24mm}\rule{0.2mm}{2.5mm}\hspace{-2.7mm} C}}
\nc{\Nat}{\mbox{\hspace{.04mm}\rule{0.2mm}{2.8mm}\hspace{-1.5mm} N}}

\nc{\spa}{\hspace{.1cm},\hspace{1 cm}}
\nc{\vs}{\vspace}
\nc{\NP}[1]{Nucl.\ Phys.\ {\bf #1}}
\nc{\PL}[1]{Phys.\ Lett.\ {\bf #1}}
\nc{\CMP}[1]{Commun.\ Math.\ Phys.\ {\bf #1}}
\nc{\PR}[1]{Phys.\ Rev.\ {\bf #1}}
\nc{\PRL}[1]{Phys.\ Rev.\ Lett.\ {\bf #1}}
\nc{\PTP}[1]{Prog.\ Theor.\ Phys.\ {\bf #1}}
\nc{\PTPS}[1]{Prog.\ Theor.\ Phys.\ Suppl.\ {\bf #1}}
\nc{\MPL}[1]{Mod.\ Phys.\ Lett.\ {\bf #1}}
\nc{\IJMP}[1]{Int.\ Jour.\ Mod.\ Phys.\ {\bf #1}}
\nc{\IM}[1]{Invent.\ Math.\ {\bf #1}}
\nc{\SJNP}[1]{Sov. J. Nucl. Phys.\ {\bf #1}}
\nc{\JHEP}[1]{J.\ High\ Energy Phys.\ {\bf #1}}

\begin{document}

\topmargin -5mm
\oddsidemargin 5mm

\begin{titlepage}
\setcounter{page}{0}
\begin{flushright}
February 2000\\
TIFR/TH/99-58
\end{flushright}

\vs{8mm}
\begin{center}
{\Large Constructing Classical and Quantum Superconformal Algebras}\\[.2cm]
{\Large on the Boundary of $AdS_3$}

\vs{8mm}
{\large J{\o}rgen Rasmussen}\footnote{e-mail address: 
rasmussj@cs.uleth.ca}
\\[.2cm]
{\em Department of Theoretical Physics, 
Tata Institute of Fundamental Research}\\
{\em Homi Bhabha Road, Colaba, Mumbai 400 005, India}\\[.1cm]
and\\[.1cm]
{\em Physics Department, University of Lethbridge}\\
{\em Lethbridge, Alberta, Canada T1K 3M4}\footnote{Address after 
10th of February 2000}

\end{center}

\vs{8mm}
\centerline{{\bf{Abstract}}}
\noindent
Motivated by recent progress on the correspondence between string theory on
anti-de Sitter space and conformal field theory, we address the question
of constructing space-time $N$ extended  superconformal algebras on 
the boundary of $AdS_3$. 
Based on a free field realization of an affine $SL(2|N/2)$ current 
superalgebra residing on the world sheet, we construct
explicitly the Virasoro generators and the $N$ supercurrents. 
$N$ is even. The resulting superconformal algebra 
has an affine $SL(N/2)\otimes U(1)$ current algebra
as an internal subalgebra. Though we do not complete the
general superalgebra, we outline the underlying construction
and present supporting evidence for its validity. Particular
attention is paid to its BRST invariance.  
In the classical limit where the free field realization may be
substituted by a differential operator realization, we discuss
further classes of generators needed in the closure of the algebra.
We find sets of half-integer spin fields, and for $N\geq6$ these include
generators of negative weights. An interesting property of the construction
is that for $N\neq2$ it treats the supercurrents in an asymmetric way. Thus,
we are witnessing a new class of superconformal algebras not
obtainable by conventional Hamiltonian reduction.
The complete classical algebra is provided in the case $N=4$ and is of
a new and asymmetric form.
\\[.4cm]
{\em PACS:} 11.25.Hf; 11.25.-w\\
{\em Keywords:} Superconformal field theory; 
$AdS$/CFT correspondence; string theory; free field realization

\end{titlepage}
\newpage
\renewcommand{\thefootnote}{\arabic{footnote}}
\setcounter{footnote}{0}

\section{Introduction}

Recently, Maldacena proposed a duality between type IIB string 
theory and (super-) conformal field theory (CFT)
on the boundary of anti-de Sitter space ($AdS$) \cite{Mal},
further elaborated on in Refs. \cite{GKP,Wit}. This remarkable conjecture 
has induced a tremendous activity in theoretical high energy physics.
In the cases involving $AdS_3$, see e.g. Refs. 
\cite{MS,deB,GKS,KLL,deBORT,YZ,KN} and the review \cite{AGMOO}, 
the corresponding space-time CFT
is two-dimensional and thus possesses many well known properties.

The boundary of $AdS_3$ enjoys conformal symmetry, as recognized 
by Brown and Henneaux \cite{BH}.
In the work \cite{GKS}, Giveon {\em et al} have constructed explicitly the 
generators of the space-time conformal algebra from the string theory on 
$AdS_3$. The construction starts from the world sheet $SL(2)$ current 
algebra. In terms of the Wakimoto free field realization \cite{Wak} of that, 
they have provided a general expression for the Virasoro generators 
and computed the central charge. 

In Ref. \cite{Ito1} Ito has succeeded in constructing superconformal algebras
(SCAs) on the boundary of $AdS_3$ and again the construction starts from a
world sheet current algebra. However, in order to obtain an extended
conformal symmetry in space-time one needs to consider higher world sheet
current (super-)algebras than $SL(2)$. Thus, Ito has found that 
the appropriate Lie superalgebras leading to $N=1$, 2 and 4 superconformal
algebras are $osp(1|2)$, $sl(2|1)$ and $sl(2|2)$, respectively.
The explicit constructions are based on generalized Wakimoto
free field realizations of the associated
affine current superalgebras \cite{BO,Ito2,Ras}.
A related approach to construct $N=1$, 2 and 4 SCAs 
is discussed in \cite{And} in which also one- and two-point functions and
some unitary representations are considered.
In Ref. \cite{YIS} space-time $N=3$ superconformal theories are
studied.

The objective of the present paper is to take a first
step in the direction of 
classifying the SCAs that may be induced by
string theory on $AdS_3$, thus generalizing the work by Ito \cite{Ito1}.
A main property of an appropriate world sheet current (super-)algebra 
is that the bosonic part may be decomposed as $G=SL(2)\otimes G'$.
This is necessary as we want the free fields in the Wakimoto realization
of the embedded $SL(2)$ to be considered as coordinates on $AdS_3$.
As discussed in Ref. \cite{GKS}, a purely bosonic world sheet algebra
with such a decomposition leads immediately to an affine Lie
algebra on the boundary of $AdS_3$. The present paper is devoted to
discussing the class of {\em superconformal} algebras that may be constructed
starting from affine $SL(2|N/2)$ current superalgebras having as bosonic
part $SL(2)\otimes SL(N/2)\otimes U(1)$. By construction, $N$ is even.

We construct explicitly the Virasoro generators, the $N$ supercurrents,
and the generators of an internal $SL(N/2)\otimes U(1)$ Kac-Moody
algebra.
For $N=2$, the $SL(N/2)$ and $U(1)$ current algebras collapse to a single
$U(1)$ current algebra. This conventional $N=2$ superconformal
algebra has already been obtained by Ito \cite{Ito1}. For higher $N$ we turn 
to a classical limit in which the generators may be substituted by
first order linear differential operators. The resulting classical
SCAs are center-less and are shown to include classes of primary 
generators of half-integer weights which are all smaller than 2.
Some weights are negative for $N\geq6$. SCAs based on free field
realizations and with generically non-vanishing central charges 
are denoted {\em quantum} SCAs as opposed to such {\em classical}
SCAs. Thus, our use of the notion quantum is not in the
quantum group sense of $q$-deformations.

A new and important property of the construction is that for $N\neq2$
it treats the
supercurrents asymmetrically. This is illustrated in the case $N=4$
where the classical SCA is completed and found to be of a new and
asymmetric form. Thus, it is not included in the standard classification
of $N=4$ SCAs \cite{Aetal,Sch,STV,AK,Ali}. 
In particular, it deviates essentially from the small $N=4$ SCA
which has otherwise been announced to be the result \cite{Ito1}
of a construction similar to the one employed in the present paper.
This is argued not to be the correct result. 
The full quantum $N=4$ SCA with generic
central charge will be presented elsewhere \cite{Ras2}.
There we shall also show that the {\em small} $N=4$ SCA may be obtained
by replacing the original world sheet $SL(2|2)$ current superalgebra
by the related $SL(2|2)/U(1)$ current superalgebra.

A complete classification along the lines indicated
is reached when the SCAs induced by any world sheet current superalgebra with
$SL(2)\otimes G'$ decomposable bosonic part have been constructed.
We anticipate that the techniques employed in the present paper may be
enhanced to cover the general case and hope to come back elsewhere
with a discussion on this generalization.

As pointed out in Ref. \cite{GKS}, BRST invariance of the construction
of the space-time conformal algebra is equivalent to requiring the Virasoro
generators to be primary fields of weight one with respect to the world sheet
energy-momentum tensor, ensuring that the integrated fields commute with
the world sheet Virasoro algebra. This carries over to the superconformal 
case, and we shall verify that the generators of our 
SCAs meet the requirement of being primary of weight one
with respect to the world sheet current superalgebra Sugawara tensor.  

The algebras constructed in Ref. \cite{And} are simpler than the 
ones by Ito \cite{Ito1} as they are based on smaller Lie superalgebras. For 
example, the $N=4$ SCA is constructed from an $sl(2|1)$ Lie
superalgebra. However, the central charges are essentially fixed, and
the algebras are in general not ensured to be BRST invariant\footnote{We
thank O. Andreev for pointing out that the $N=4$ SCA
in Ref. \cite{And} is nevertheless BRST invariant.}.

The remaining part of this paper is organized as follows.
In Section 2 we review the construction of the Virasoro algebra and the
immediate extension to an affine Lie algebra \cite{GKS}.

In Section 3 we introduce our notation for Lie superalgebras and
their associated current superalgebras, and review the free field 
realizations of the latter obtained in Ref. \cite{Ras}.

In Section 4 we provide our explicit construction of the supercurrents,
the Virasoro generators, and the generators of the internal
$SL(N/2)\otimes U(1)$ Kac-Moody algebra.

In Section 5 BRST invariance of the construction is addressed.

In Section 6 we discuss the classical $N$ extended SCAs and write down
the explicit result for $N=4$.

Section 7 contains concluding remarks, while details on the Lie
superalgebra $sl(2|M)$ are given in Appendix A.

\section{Virasoro Algebra}

The standard Wakimoto free field realization of the affine
$SL(2)$ current algebra with level $\kn$ \cite{Wak} is
\bea
 E&=&\beta\nn
 H&=&-2\g\beta+\sqrt{\kn+2}\pa\varphi\nn
 F&=&-\g^2\beta+\sqrt{\kn+2}\g\pa\varphi+\kn\pa\g
\label{sl2}
\eea
Here and throughout the paper, normal ordering is implicit.
The operator product expansions (OPEs) of the ghost fields $\beta,\g$ and
the bosonic scalar field $\var$ are 
\ben
 \beta(z)\g(w)=\frac{1}{z-w}\spa\var(z)\var(w)=2\ln(z-w)
\een
where regular terms have been omitted.
In Ref. \cite{GKS} it is shown that the world sheet $SL(2)$ current algebra
with level $\kn$ induces the Virasoro algebra 
\ben
 [L_n,L_m]=(n-m)L_{n+m}+\frac{c}{12}(n^3-n)\delta_{n+m,0}
\label{Vir}
\een
on the boundary of $AdS_3$. The generators are given by
\bea
 L_n&=&\oint\dz{\cal L}_n(z)\nn
 {\cal L}_n&=&a_+(n)\g^{n+1}E+a_3(n)\g^nH+a_-(n)\g^{n-1}F
\label{virsl2}
\eea
with constants\footnote{Note that the conventions used
here differ slightly from the ones used in Ref. \cite{GKS}, which is also 
reflected in a sign discrepancy in the central charge (\ref{c}).}
\bea
 a_+(n)&=&\hf(1-n)n\nn
 a_3(n)&=&\hf(1-n)(1+n)\nn
 a_-(n)&=&\hf n(1+n)
\label{al}
\eea
The central charge is found to be 
\ben
 c=-6\kn p
\label{c}
\een
where $p$ is the integer winding number
\ben
 p=\oint\dz \frac{\pa\g}{\g}
\label{p}
\een 

BRST invariance requires ${\cal L}_n$ to be conformal primary of weight
1 with respect to the world sheet Virasoro generator
\ben
 T=\pa\g\beta+\hf\pa\var\cdot\pa\var-\frac{1}{2\sqrt{\kn+2}}\pa^2\var
\een
This property is readily verified.

The Virasoro algebra is immediately extended to an affine Lie algebra if the
world sheet affine $SL(2)$ current algebra is replaced by an affine 
$G=SL(2)\otimes G'$ current algebra, where $G'$ is a Lie group.
Indeed, let $J_a$ denote the currents of the affine $G'$ current algebra
with central extension $k'$, and define the generators
\ben
 I_{a;n}=\oint\dz\g^n(z)J_a(z)
\een
{}From the defining OPE
\ben
 J_a(z)J_b(w)=\frac{\kappa_{a,b}k'}{(z-w)^2}+\frac{\fabc J_c(w)}{z-w}
\een
where $\kappa_{a,b}$ and $\fabc$ are the Cartan-Killing form and the structure
constants, respectively, of the underlying Lie algebra ${\bf g'}$, one finds
\bea
 \left[L_n,I_{b;m}\right]&=&-mI_{b;n+m}\nn
 \left[I_{a;n},I_{b;m}\right]&=&\fabc I_{c;n+m}+npk'\kappa_{a,b}\delta_{n+m,0}
\label{II}
\eea
Summation over ``properly'' repeated indices is implicit.
The central extension $k'p$ of the space-time affine Lie algebra 
generated by $\left\{ I_{a;n}\right\}$ is seen to be the one, $k'$, 
of the world sheet $G'$ current algebra multiplied by the winding
number $p$ of the embedded $sl(2)$ subalgebra. As the currents $J_a$
are spin one primary fields, the construction is
readily seen to be BRST invariant due to the decomposition $SL(2)\otimes G'$.

\section{Affine Current Superalgebra}

\subsection{Lie Superalgebra}

Let {\bf g}$^0$ and {\bf g}$^1$ denote the even and odd parts, respectively,
of the Lie superalgebra {\bf g} of rank $r$, see Ref. \cite{Kac} and
references therein. $\D=\D^0\cup\D^1$ is the set of roots $\al$ of {\bf g}
where $\D^0$ ($\D^1$) is the set of even (odd) roots.
The set of positive roots $\al>0$ is $\D_+=\D^0_+\cup\D^1_+$.
A choice of simple roots is written $\left\{\al_i\right\}_{i=1,...,r}$.
A distinguished representation is characterized by exactly one simple
root being odd. Related to the triangular decomposition
\ben
 {\bf g}={\bf g}_-\oplus{\bf h}\oplus{\bf g}_+
\een
the raising and lowering operators are denoted $E_\al,J_\al\in{\bf g}_+$ and
$F_\al,J_{-\al}\in{\bf g}_-$, respectively, with $\al\in\D_+$, while
$H_i,J_i\in{\bf h}$ are the Cartan generators. Generic Lie superalgebra
elements are denoted $J_a$ and satisfy
\ben
 \left[J_a,J_b\right\}=\fabc J_c 
\een
where $\left[\cdot,\cdot\right\}$ is an anti-commutator if both arguments
are fermionic, and otherwise a commutator.
The Jacobi identities read
\ben
 \left[J_a,\left[J_b,J_c\right\}\right\}=\left[\left[J_a,J_b\right\},
  J_c\right\}+(-1)^{p(J_a)p(J_b)}\left[J_b,\left[J_a,J_c\right\}\right\}
\een
where the parity $p(J_a)$ is 1 (0) for $J_a$ an odd (even) generator.
The Cartan-Killing form $\kappa_{a,b}$ 
\ben
 2\hn\kappa_{a,b}=\mbox{str}(\mbox{ad}_{J_a}\mbox{ad}_{J_b})
\een
and the Cartan matrix $A_{ij}=\al_j(H_i)$
are related as $\kappa_{i,j}=A_{ij}\kappa_{\al_j,-\al_j}$. $\hn$ is the dual
Coxeter number. The Weyl vector 
\bea
 &\rho=\rho^0-\rho^1&\nn
 \rho^0=\hf\sum_{\al\in\D^0_+}\al&,&\rho^1=\hf\sum_{\al\in\D^1_+}\al 
\eea
satisfies $\rho\cdot\al_i=\al_i^2/2$.

For each positive even or odd root $\al>0$ we introduce a super-triangular 
coordinate denoted by $x^\al$ or $\theta^\al$, respectively, where
$\theta^\al$ is Grassmann odd. In terms of the matrix 
\ben
 C_a^b(x,\theta)=-\sum_{\al\in\D^0_+}x^\al{f_{\al,a}}^b
  -\sum_{\al\in\D^1_+}\theta^\al{f_{\al,a}}^b
\label{cadj}
\een
one may then realize the Lie superalgebra in terms of differential operators
\cite{Ras}
\ben
 J_a(x,\theta,\pa,\La)=\sum_{\al>0}V_a^\al(x,\theta)\pa_\al+\sum_{j=1}^r
  P_a^j(x,\theta)\La_j
\label{diff}
\een
where $\La$ is the weight of the representation, and $\La_j$ are the
labels defined by
\ben
 H_j\ket{\La}=\La(H_j)\ket{\La}=\La_j\ket{\La}
\een 
$\pa_\al$ is differentiation with respect to $x^\al$ or $\theta^\al$ 
depending on the
parity of $\al$, whereas $V$ and $P$ are finite dimensional polynomials:
\bea
  V_\al^{\al'}(x,\theta)&=&\left[B(C(x,\theta))\right]_\al^{\al'}\nn
  V_i^{\al'}(x,\theta)&=&-\left[C(x,\theta)\right]_i^{\al'}\nn
  V_{-\al}^{\al'}(x,\theta)&=&\sum_{\al''>0}\left[e^{-C(x,\theta)}
  \right]_{-\al}^{\al''}\left[B(-C(x,\theta))\right]_{\al''}^{\al'}\nn
  P_\al^j(x,\theta)&=&0\nn
  P_i^j(x,\theta)&=&\delta_i^j\nn
  P_{-\al}^j(x,\theta)&=&\left[e^{-C(x,\theta)}\right]_{-\al}^j
\label{pol}
\eea
$B(u)$ is the generating function for the Bernoulli numbers $B_n$
\bea
 B(u)&=&\frac{u}{e^u-1}=\sum_{n\geq0}\frac{B_n}{n!}u^n\nn
 (B(u))^{-1}&=&\frac{e^u-1}{u}=\sum_{n\geq0}\frac{1}{(n+1)!}u^n
\label{Ber}
\eea
The formal power series expansions (\ref{pol}) all truncate and
become polynomials due to the nilpotency of the matrix $C$ (\ref{cadj}).
For later use, let us also introduce the notation $\vpp$ for the
first of the polynomials in (\ref{pol})
\ben
  \vpp(x,\theta)={[B(C(x,\theta))]_+}^+=B(\cpp(x,\theta))
\een
where $\cpp$ is the submatrix of $C$ (\ref{cadj}) with both row and column
indices positive (even or odd) roots. $\vpp$ is immediately seen to
be invertible
\ben
 (\vpp(x,\theta))^{-1}=(B(\cpp(x,\theta)))^{-1}=
  \sum_{n\geq0}\frac{1}{(n+1)!}(\cpp(x,\theta))^n
\label{vminus}
\een

Most Lie superalgebras with even subalgebra ${\bf g}^0=sl(2)\oplus{\bf g'}$
have the property that the embedding of $sl(2)$ in ${\bf g}$
carried by ${\bf g}^1$ is a spin 1/2 representation\footnote{This is true
for all basic Lie superalgebras with even subalgebra  
${\bf g}^0=sl(2)\oplus{\bf g'}$ except $osp(3|2M)$ where the embedding 
is a spin 1 representation, see e.g. \cite{IMP}.}. This means that
the space of odd roots may be divided into two parts
\ben
 \D^1=\D^{1-}\cup\D^{1+}
\label{decomp}
\een
where the roots $\al^\pm\in\D^{1\pm}$ are characterized by 
\ben
 \frac{\al_{sl(2)}\cdot\al^\pm}{\al_{sl(2)}^2}=\pm\hf
\een
and we have the correspondence
\ben
 \D^{1+}=\al_{sl(2)}+\D^{1-}
\een
$\al_{sl(2)}$ is the positive root associated to
the embedded $sl(2)$. In particular, the division (\ref{decomp})
is present in the case of our main interest, namely the Lie superalgebra 
$sl(2|M)$ which is considered in Section 4 and further in Appendix A.

\subsection{Free Field Realization}

Associated to a Lie superalgebra is an affine Lie superalgebra 
characterized by the central extension $k$, and associated to an affine Lie
superalgebra is an affine current superalgebra whose generators are conformal
spin one primary fields and have the mutual operator product
expansions\footnote{We note that the extension of the Virasoro algebra
to include an affine Lie algebra (\ref{II}) discussed in Section 2 may
readily be generalized to an affine Lie superalgebra simply by
substituting the Lie group $G'$ with a Lie supergroup; the only change
being that the commutator $[I_{a;n},I_{b;m}]$ becomes an anti-commutator
for $J_a$ and $J_b$ both fermionic.} 
\ben
 J_a(z)J_b(w)=\frac{\kappa_{a,b}k}{(z-w)^2}+\frac{{f_{a,b}}^c J_c(w)}{z-w}
\label{JaJb}
\een
We use the same notation $J,E,F,H$ for the currents as for the algebra 
generators. Hopefully, this will not lead to misunderstandings. 
The associated Sugawara construction 
\ben
 T=\frac{1}{2(k+\hn)}\kappa^{a,b}J_aJ_b
\een
generates the Virasoro algebra with central charge
\ben
 c=\frac{k\ \mbox{sdim}(\mbox{{\bf g}})}{k+\hn}
\label{csug}
\een

The standard free field construction \cite{BO,Ito2,Ras}
consists in introducing for every positive {\em even}
root $\al\in\D^0_+$, a pair of free {\em bosonic} ghost
fields ($\ba,\ga$) of conformal weights (1,0) satisfying the OPE
\ben
 \ba(z)\g^{\al'}(w)=\frac{{\delta_{\al}}^{\al'}}{z-w}
\een
The corresponding energy-momentum tensor is
\ben
 T_{\beta\g}=\sum_{\al\in\D_+^0}\pa\ga\ba
\label{Tbg}
\een
with central charge
\ben
 c_{\beta\g}=2|\D_+^0|=\mbox{dim}(\mbox{{\bf g}}^0)-r
\een
For every positive {\em odd} root $\al\in\D^1_+$ one introduces a pair of free
{\em fermionic} ghost fields ($b_\al,c^\al$) of conformal weights
(1,0) satisfying the OPE
\ben
 b_\al(z)c^{\al'}(w)=\frac{{\delta_{\al}}^{\al'}}{z-w}
\een
The corresponding energy-momentum tensor is
\ben
 T_{bc}=\sum_{\al\in\D_+^1}\pa c^\al b_\al
\label{Tbc}
\een
with central charge
\ben
 c_{bc}=-2|\D_+^1|=-\mbox{dim}(\mbox{{\bf g}}^1)
\een
For every Cartan index $i=1,...,r$ one introduces a free scalar boson $\var_i$
with contraction
\ben
 \var_i(z)\var_j(w)=\kappa_{i,j}\ln(z-w)
\een
The corresponding energy-momentum tensor 
\ben
 T_\var=\hf\pa\var\cdot\pa\var-\frac{1}{\sqrt{k+\hn}}\rho\cdot\pa^2\var
\een
has central charge
\ben
 c_\var=r-\frac{\hn \mbox{sdim}(\mbox{{\bf g}})}{k+\hn}
\label{cvar}
\een
where the super-dimension sdim$(\mbox{{\bf g}})$ of the Lie superalgebra 
{\bf g} is defined as the difference dim({\bf g}$^0$) $-$ dim({\bf g}$^1$).
In obtaining (\ref{cvar}) we have used Freudenthal-de Vries (super-)strange
formula 
\ben
 \rho^2=\frac{\hn}{12}\ \mbox{sdim}(\mbox{{\bf g}})
\een
The total free field realization of the Sugawara energy-momentum tensor 
is $T=T_{\beta\g}+T_{bc}+T_\var$ and has indeed central charge (\ref{csug}).

The generalized Wakimoto free field realization of the affine current
superalgebra is obtained by the substitution
\bea
 \pa_{x^\al}\rightarrow\beta_\al(z)&\spa&x^\al\rightarrow\g^\al(z)
  \spa\La_i\rightarrow\sqrt{k+\hn}\pa\var_i(z)\nn
 \pa_{\theta^\al}\rightarrow b_\al(z)&\spa&\theta^\al\rightarrow c^\al(z)
\label{subst}
\eea
in the differential operator realization 
$\left\{J_a(x,\theta,\pa,\La)\right\}$ (\ref{diff}), (\ref{pol}), 
and a subsequent addition of anomalous terms linear in $\pa\g$ or $\pa c$:
\bea
 J_a(z)&=&\sum_{\al\in\D^0_+}V_a^\al(\g(z),c(z))\beta_\al(z)
  +\sum_{\al\in\D^1_+}V_a^\al(\g(z),c(z))b_\al(z)\nn
 &+&\sqrt{k+\hn}\sum_{j=1}^rP_a^j(\g(z),c(z))\pa\var_j(z)
  +J^{\mbox{\tiny{anom}}}_a(\g(z),c(z),\pa\g(z),\pa c(z))
\label{subst2}
\eea
Anomalous terms are only added to the lowering generators $F_\al(z)$
\ben
 J^{\mbox{\tiny{anom}}}_a(\g(z),c(z),\pa\g(z),\pa c(z))= 
  \left\{\begin{array}{l}0\ \ \ \mbox{for}\ \ a=i,\al>0\\   {}\\
  \sum_{\al'\in\D^0_+}\pa\g^{\al'}(z)F_{\al,\al'}(\g(z),c(z))\\ 
  +\sum_{\al'\in\D^1_+}\pa c^{\al'}(z)F_{\al,\al'}(\g(z),c(z))\ \ \ \mbox{for}
  \ \ a=\al<0  \end{array} \right.
\een
and are given by
\bea
 F_{\al\in\D^0_+,\al'}(\g,c)&=&k\sum_{\mu\in\D_+}
  \left[(\vpp(\g,c))^{-1}\right]_{\al'}^\mu\kappa_{\mu,-\al}\nn
 &+&\sum_{\mu,\sigma\in\D^0_+,\nu\in\D_+}
  \left[(\vpp(\g,c))^{-1}\right]_{\al'}^\mu
  \pa_\sigma V_\mu^\nu(\g,c)\pa_\nu V_{-\al}^\sigma(\g,c)\nn
 &-&\sum_{\mu\in\D^0_+,\sigma\in\D^1_+,\nu\in\D_+}
  \left[(\vpp(\g,c))^{-1}\right]_{\al'}^\mu\pa_\sigma V_\mu^\nu(\g,c)
  \pa_\nu V_{-\al}^\sigma(\g,c)\nn
 &+&\sum_{\mu\in\D^1_+,\sigma,\nu\in\D_+}
  \left[(\vpp(\g,c))^{-1}\right]_{\al'}^\mu\pa_\sigma
  V_\mu^\nu(\g,c)\pa_\nu V_{-\al}^\sigma(\g,c)\nn
 F_{\al\in\D^1_+,\al'}(\g,c)&=&k\sum_{\mu\in\D_+}
  \left[(\vpp(\g,c))^{-1}\right]_{\al'}^\mu\kappa_{\mu,-\al}\nn
 &+&\sum_{\mu,\in\D^0_+,\sigma,\nu\in\D_+}
  \left[(\vpp(\g,c))^{-1}\right]_{\al'}^\mu
  \pa_\sigma V_\mu^\nu(\g,c)\pa_\nu V_{-\al}^\sigma(\g,c)\nn
 &+&\sum_{\mu\in\D^1_+,\sigma\in\D^0_+,\nu\in\D_+}
  \left[(\vpp(\g,c))^{-1}\right]_{\al'}^\mu\pa_\sigma V_\mu^\nu(\g,c)
  \pa_\nu V_{-\al}^\sigma(\g,c)\nn
 &-&\sum_{\mu,\sigma\in\D^1_+,\nu\in\D_+}
  \left[(\vpp(\g,c))^{-1}\right]_{\al'}^\mu\pa_\sigma
  V_\mu^\nu(\g,c)\pa_\nu V_{-\al}^\sigma(\g,c)
\eea
In particular, for $\al$ a simple root the anomalous term $F_{\al,\al'}$ 
is a constant independent of $\g$ and $c$:
\ben
 F_{\al_i,\al'}(\g,c)
  =\hf\delta_{\al_i,\al'}\left((2k+\hn)\kappa_{\al_i,-\al_i}-A_{ii}\right)
\label{anomsimple}
\een
This concludes the explicit free field realization of general 
affine current superalgebras obtained in Ref. \cite{Ras}, where the 
polynomials $V$, $P$ and $(\vpp)^{-1}$ are given in (\ref{pol}) and
(\ref{vminus}).

\section{Generators of the Superconformal Algebra}

In this section we shall construct the Virasoro generators and
the supercurrents of the SCA in space-time
that may be induced by the affine $SL(2|N/2)$ current superalgebra on
the world sheet. Some steps towards constructing the complete set
of SCA generators are also taken.
For simplicity, we consider the underlying Lie
superalgebra $sl(2|N/2)$ in the distinguished representation, see Appendix A.
Let us introduce the abbreviations
$\g=\g^{\al_1}$, $E_1=E_{\al_1}$ etc for the objects related to the
embedded $sl(2)$. Hopefully, no misunderstandings will arise, as we are also
using $\g$ to represent a general bosonic ghost field argument in the
polynomials $V$ and $P$, though in general we will leave out the arguments. 

Using the explicit polynomials
\bea
 V_{\al_1}^{\al_1}(\g,c)&=&1\nn
 V_1^{\al_1}(\g,c)&=&-2\g\nn
 V_{-\al_1}^{\al_1}(\g,c)&=&-\g^2
\label{V1}
\eea
it is straightforward to verify that the Virasoro algebra (\ref{Vir})
is generated by
\bea
 L_n&=&\oint\dz{\cal L}_n(z)\nn
 {\cal L}_n&=&a_+(n)\g^{n+1}E_1+a_3(n)\g^nH_1+a_-(n)\g^{n-1}F_1
\label{vir}
\eea
and has central charge
\ben
 c=-6k^\vee_1p_1
\een
$p_1$ is the winding number (\ref{p}) for the ghost field
$\g^{\al_1}=\g$, and $k^\vee_1=\kappa_{\al_1,-\al_1}k$ is the level of
the embedded $sl(2)$ or the level in the direction $\al_1$. 

As a preparation for 
constructing the algebra generators, let us introduce the generators
\ben
 J_{a;n}=\oint\dz\g^n(z)J_a(z)
\een
and consider the OPE
\bea
 {\cal L}_n(z)J_a(w)&=&\frac{1}{z-w}\left\{\left(a_+(n)\g^{n+1}{f_{\al_1,a}}^c
  +a_3(n)\g^n{f_{1,a}}^c+a_-(n)\g^{n-1}{f_{-\al_1,a}}^c\right)J_c\right.\nn
 &&-na_3(n)\g^{n-2}\left.V_a^{\al_1}\left(\g^2E_1+\g H_1
  -F_1\right)\right\}\nn
 &-&\frac{1}{(z-w)^2}na_3(n)\g^{n-2}(z)\left(\g^2(z)V_{\al_1}^\nu(z)
  +\g(z)V_1^\nu(z)-V_{-\al_1}^\nu(z)\right)\pa_\nu V_a^{\al_1}(w)\nn
\label{LJ}
\eea
Here and in the following summations over repeated indices
are implicit. Summations over
root indices are meant to be over all {\em positive} roots if not otherwise 
indicated. Actually, this restriction is not necessary as the 
super-triangular
coordinates are defined for positive roots only, i.e. $\pa_\nu$ exists
only for $\nu>0$. Now, due to the structure of the root space 
(see Appendix A) we immediately obtain
\ben
 \left[L_n,J_{\al;0}\right]=0\ \ \ \mbox{for} \ \ \al\in\D^0\setminus
  \left\{\pm\al_1\right\}
\label{LJdelta}
\een
Likewise, it follows that
\bea
 \left[L_n,J_{\al^-;0}\right]&=&\hf(n-1)\g^n\left((n+1)J_{\al^-;n}
  -n\g J_{\al^+;n+1}\right)\nn
 \left[L_n,J_{\al^+;0}\right]&=&\hf(n+1)\g^{n-1}\left(nJ_{\al^-;n-1}
  -(n-1)\g J_{\al^+;n}\right)
\label{LJalpha}
\eea
The idea is to use the right hand sides in the construction of the
supercurrents. To this end let us consider the general setting where
a primary field $\Phi$ of weight $h$ 
\ben
 \left[L_n,\Phi_{m+\eta}\right]=\left((h-1)n-m-\eta\right)\Phi_{n+m+\eta}
\een
may be obtained as a commutator of the form
\ben
 \left[L_m,\oint B\right]=b(m;\eta;h)\Phi_{m+\eta}
\een
$\eta$ is a possible non-integer shift in the modes, whereas 
$b$ is some function. From the Jacobi identities this function
satisfies the recursion relation
\ben
 (n-m)b(n+m)=((h-1)n-m-\eta)b(m)-((h-1)m-n-\eta)b(n)
\een
allowing the simple solution
\ben
 b(m)=b_0+b_1m\spa (1-h)b_0=\eta b_1
\een
This is precisely of the form we have encountered in (\ref{LJalpha}).
Since we want to construct supercurrents of weight 3/2 we should
choose $\eta\in\Z+\hf$, and we define the generators
\bea
 G_{\al^-;n+1/2}&=&\oint\dz{\cal G}_{\al^-;n+1/2}(z)\nn
 {\cal G}_{\al^-;n+1/2}&=&(n+1)\g^nJ_{\al^-}-n\g^{n+1}J_{\al^+}
\eea
Of course, it still remains to verify that the supercurrents $G$  
are indeed primary:
\ben
 [L_n,G_{\al^-;m+1/2}]=\left(\hf n-m-\hf\right)G_{\al^-;n+m+1/2}
\label{LG}
\een
However, that follows immediately from the
simple computation of the OPE ${\cal L}_n{\cal G}_{\al^-;m+1/2}$,
and we have thus constructed half of the supercurrents.
Note that both commutators in (\ref{LJalpha}) lead to the same
supercurrent as we have
\ben
 G_{\al^-;n+1/2}=\left\{\begin{array}{l}\frac{2}{n-1}\left[L_n,J_{\al^-;0}
   \right]\\ \\
  \frac{2}{n+2}\left[L_{n+1},J_{\al^+;0}\right]\end{array}\right.
\label{GLJ}
\een
This means that $G$ may be represented as a commutator for all integer
modes $n$.

Let us now turn to the construction of the supercurrents $\Gb$.
For $a=-\al^\pm\in\D_-^{1\pm}$ it follows from (\ref{LJ}) that a situation 
like (\ref{GLJ}) occurs provided
\ben
 V_{-\al^+}^{\al_1}=-\g V_{-\al^-}^{\al_1}
\label{VpVm}
\een
This important relation is proven in Appendix A where also some identities
involving $\Vm$ are derived. We are led to define the supercurrents
$\Gb$ by
\bea
 \Gb_{-\al^-;n-1/2}&=&\oint\dz\Gbc_{-\al^-;n-1/2}(z)\nn
 \Gbc_{-\al^-;n-1/2}&=&(n-1)\g^nJ_{-\al^-}+n\g^{n-1}J_{-\al^+}\nn
 &-&n(n-1)\g^{n-2}\Vm\left(\g^2E_1+\g H_1-F_1\right)\nn
 &+&n(n-1)\g^{n-2}\left(\g^2V_{\al_1}^\nu+\g V_1^\nu
  -V_{-\al_1}^\nu\right)\pa_\nu\pa_\sigma\Vm\pa\g^\sigma\nn
 &=&(n-1)\g^nJ_{-\al^-}+n\g^{n-1}J_{-\al^+}\nn
 &-&n(n-1)\g^{n-2}\left(\G^\nu_{-\al_1} b_\nu-(k-1+\hn/2)\pa\g-\pa\g^\sigma
  \G^\nu_{-\al_1}\pa_\nu\pa_\sigma\right)\Vm
\label{Gb}
\eea
where 
\ben
 \G^\nu_{-\al_1}=\g^2V_{\al_1}^\nu+\g V_1^\nu-V_{-\al_1}^\nu
\label{Gamma}
\een
As indicated in (\ref{Gb}), one may show that the upper root index $\nu$ is 
always an odd (and positive) root. The analogue to (\ref{GLJ}) reads
\ben
 \Gb_{-\al^-;n-1/2}=\left\{\begin{array}{l}\frac{-2}{n+1}\left[L_n,
  J_{-\al^-;0}\right]\\ \\
  \frac{2}{n-2}\left[L_{n-1},J_{-\al^+;0}\right]\end{array}\right.
\label{GLJm}
\een
As in the case of the supercurrents $G$, there is a free overall scaling.
However, in order to produce the conventional prefactor of plus one 
multiplying the
Virasoro generator in the anti-commutator $\left\{G,\Gb\right\}$
(see (\ref{GGb})), the {\em relative} factor is fixed\footnote{ A
more commonly used convention is a prefactor of plus two. However,
we have found it natural to define $G$ and $\Gb$ without
introducing any powers of $\sqrt{2}$. To comply with the standard
convention is straightforward, though.}.

Before proving that $\Gb$ is a primary field of weight 3/2
\ben
 [L_n,\Gb_{-\al^-;m-1/2}]=\left(\hf n-m+\hf\right)\Gb_{-\al^-;n+m-1/2}
\label{LGb}
\een
let us observe the following property of the construction.
Consider the Jacobi identity
\ben
 [J_{\delta;0},[L_n,J_{\pm\al^-;0}]]+[L_n,[J_{\pm\al^-;0},J_{\delta;0}]]
  =[J_{\pm\al^-;0},[L_n,J_{\delta;0}]]=0
\een
where $\delta\in\D^0\setminus\left\{\pm\al_1\right\}$ is any
even (positive or negative) root different from $\pm\al_1$. {}From
the construction of $G$ and $\Gb$ it follows that\footnote{Subtleties
for $n=\pm1,\pm2$ are immediately resolved by the alternative
commutator representations (\ref{GLJ}) and (\ref{GLJm}).}
\bea
 G_{\beta^-;n+1/2}&=&-\left[J_{\beta^--\al^-;0},G_{\al^-;n+1/2}\right],
  \ \ \ \mbox{for}\ \ \beta^--\al^-\in\D^0\setminus\left\{\pm\al_1
  \right\}\nn
 \Gb_{-\beta^-;n-1/2}&=&\left[J_{\al^--\beta^-;0},\Gb_{-\al^-;n-1/2}
  \right],\ \ \ \mbox{for}\ \ \beta^--\al^-\in\D^0\setminus\left\{
  \pm\al_1\right\}
\label{JG}
\eea
Here we have used that
\ben
 {f_{\al^-,\beta^--\al^-}}^{\beta^-}=-{f_{-\al^-,\al^--\beta^-}}^{-\beta^-}
  =1,\ \ \ \mbox{for}\ \ \beta^--\al^-\in\D^0\setminus\left\{
  \pm\al_1\right\}
\een
Besides providing information on the underlying algebraic structure of 
our construction, the translational property (\ref{JG}) may be used to
reduce considerations for general supercurrents to similar ones for the
supercurrents $G_{\al_2;n+1/2}$ and in particular $\Gb_{-\al_2;n-1/2}$.
$\al_2$ is the only fermionic simple root, see Appendix A.
Thus, as a first application we shall prove that $\Gb$ is primary.
{}From the Jacobi identities we find
\ben
 [L_n,\Gb_{-\al^-;m-1/2}]=[J_{\al_2-\al^-;0},[L_n,\Gb_{-\al_2;m-1/2}]]
\een
leaving us with the task of proving that $\Gb_{-\al_2}$ is primary.
To that end we work out
\ben
 V_{-\al_2}^{\al_1}=\G_{-\al_1}^{\al_2}=\hf\g c+C
\een
where $c$ and $C$ are the fermionic ghost fields associated to the
odd roots $\al_2$ and $\al_1+\al_2$, respectively, and the supercurrent 
becomes
\bea
 \Gbc_{-\al_2;m-1/2}&=&(m-1)\g^mJ_{-\al_2}+m\g^{m-1}J_{-\al_1-\al_2}\nn
 &-&m(m-1)\g^{m-2}V_{-\al_2}^{\al_1}\left(\g^2E_1+\g H_1-F_1-\hf\pa\g\right)
\eea
Now, one may compute the OPE ${\cal L}_n\Gbc_{-\al_2;m-1/2}$ and reduce the 
result using
\ben
 V_{\al_1}^{\al_1+\al_2}=-\hf c\spa V_1^{\al_1+\al_2}=-C\spa
  V_{-\al_1}^{\al_1+\al_2}=-\hf\g V_{-\al_2}^{\al_1}
\een
to the desired commutator
\ben
 [L_n,\Gb_{-\al_2;m-1/2}]=\left(\hf n-m+\hf\right)\Gb_{-\al_2;n+m-1/2}
\een
This concludes the proof of (\ref{LGb})
that $\Gb_{-\al^-}$ is primary of weight 3/2.

\subsection{Affine $SL(N/2)\otimes U(1)$ Current Subalgebra}
 
In order to derive the entire set of generators of the SCA, 
one should first consider the anti-commutators
$\left\{ G_{\al^-},G_{\beta^-}\right\}$, $\left\{G_{\al^-},
\Gb_{-\beta^-}\right\}$ and $\left\{\Gb_{-\al^-},\Gb_{-\beta^-}\right\}$.
It is readily seen that
\ben
 \left\{G_{\al^-;n+1/2},G_{\beta^-;m+1/2}\right\}=0
\label{GG}
\een
whereas a rather cumbersome but essentially straightforward computation 
reveals that
\ben
 \left\{G_{\al^-;n+1/2},\Gb_{-\beta^-;m-1/2}\right\}
  =\delta_{\al^-,\beta^-}L_{n+m}+(n-m+1)K_{\al^-;-\beta^-;n+m}
  +\frac{1}{6}cn(n+1)\delta_{n+m,0}\delta_{\al^-,\beta^-}
\label{GGb}
\een
where the current $K$ is defined by
\bea
 K_{\al^-;-\beta^-;n}&=&\oint\dz{\cal K}_{\al^-;-\beta^-;n}(z)\nn
 {\cal K}_{\al^-;-\beta^-;n}&=&
  n\g^{n-1}\Vmb\left(\g J_{\al^+}-J_{\al^-}\right)
  -\g^n{f_{\al^-,-\beta^-}}^cJ_c\nn
 &+&\hf\delta_{\al^-,\beta^-}\g^{n-1}\left(n\left(\g^2E_1+\g H_1-F_1\right)-
  \g H_1\right)\nn
 &+&n\g^{n-1}\left(\g V_{\al^+}^\nu-V_{\al^-}^\nu\right)\pa_\nu\pa_\sigma\Vmb
  \pa\g^\sigma
\label{K}
\eea
There are several ways of representing $K_{\al^-;-\beta^-;n}$ of which
the following two turn out to be useful
\ben
 K_{\al^-;-\beta^-;n}=\left\{\begin{array}{l}\frac{1}{n+1}\left(\left\{
  G_{\al^-;n+1/2},\Gb_{-\beta^-;-1/2}\right\}-\delta_{\al^-,\beta^-}L_n
   \right)\\ \\
  \frac{-1}{n-1}\left(\left\{G_{\al^-;1/2},\Gb_{-\beta^-;n-1/2}\right\}
  -\delta_{\al^-,\beta^-}L_n\right)  \end{array} \right.
\een
In particular, they may be used in a straightforward verification
that the current $K$ is primary of weight 1:
\ben
 [L_n,K_{\al^-;-\beta^-;m}]=-mK_{\al^-;-\beta^-;n+m}
\label{LK}
\een
In section 6 we shall provide evidence from considering the classical
counterpart, that $\left\{K_{\al^-;-\beta^-;n}\right\}$ generate
an affine $SL(N/2)\otimes U(1)$ current subalgebra.
The number of generators is accordingly 
\ben
 |\D_+^{1-}|^2=(N/2)^2=\mbox{dim}(sl(N/2))+1
\een

A novel feature of our construction is its asymmetry
in the two sets of supercurrents
$\left\{ G\right\}$ and $\left\{\Gb\right\}$, originating
in (\ref{GG}) and 
\ben
 \left\{\Gb_{-\al^-;n-1/2},\Gb_{-\beta^-;m-1/2}\right\}\neq0,\ \ \
  \mbox{for}\ \ \al^-\neq\beta^-,\ n\neq m,\ n+m\neq1
\label{GbGb}
\een
A proof at the classical level is presented in Section 6,
however it is obvious that a result as (\ref{GbGb}) at the
classical level remains true at the quantum level.
The right hand side of (\ref{GbGb}) involves new fields to be 
introduced in Section 6.

\subsection{Underlying Lie Superalgebra}

Here we shall express the underlying Lie superalgebra in terms of
selected modes of the SCA generators. {}From the Virasoro generator we
have
\ben
 E_1=-L_{-1},\ \ H_1=2L_0,\ \ F_1=L_1
\een
while the supercurrents allow us to write
\bea
 J_{\al^-}=G_{\al^-;1/2},&\ \ &J_{\al^+}=G_{\al^-;-1/2}\nn
 J_{-\al^-}=-\Gb_{-\al^-;-1/2},&\ \ &J_{-\al^+}=\Gb_{-\al^-;1/2}
\eea
As we have
\bea
 K_{\al^-;-\beta^-;0}&=&-\hf\delta_{\al^-,\beta^-}H_1-{f_{\al^-,-\beta^-}}^c
  J_c\nn
 \left\{J_{\eps_2-\delta_u},J_{-(\eps_2-\delta_v)}\right\}&=&
  \delta_{u,v}\left(2H_2-\sum_{i=u'}^uH_{u'+1}\right)+J_{\delta_v-\delta_u}
\eea
where $J_{\delta_v-\delta_u}$ is defined only for $v\neq u$ (see Appendix A),
we find that the remaining $(N/2)^2$ Lie superalgebra generators are given by
\bea
 J_{\delta_v-\delta_u}&=&-K_{\eps_2-\delta_u;-(\eps_2-\delta_v);0},\ \ 
  \mbox{for}\ u\neq v\nn
 H_2&=&-K_{\al_2;-\al_2;0}-L_0\nn
 H_i&=&K_{\eps_2-\delta_{i-1};-(\eps_2-\delta_{i-1});0}-
  K_{\eps_2-\delta_{i-2};-(\eps_2-\delta_{i-2});0},\ \ \mbox{for}\ 
  i=3,...,N/2+1
\eea
 
\subsection{$N=2$ Superconformal Algebra}

For $N=2$ the only positive 
$\al^-$-root is $\al_2$ (i.e. $\D_+^{1-}={\{}\al_2{\}}$)
and we have the 4 generators
\bea
 {\cal L}_n&=&a_+(n)\g^{n+1}E_1+a_3(n)\g^nH_1+a_-(n)\g^{n-1}F_1\nn
 {\cal G}_{n+1/2}&=&(n+1)\g^nJ_{\al_2}-n\g^{n+1}J_\theta\nn
 \overline{{\cal G}}_{n-1/2}&=&(n-1)\g^nJ_{-\al_2}+n\g^{n-1}J_{-\theta}\nn
 &-&n(n-1)\g^{n-2}V_{-\al_2}^{\al_1}\left(\g^2E_1+\g H_1-F_1-\hf\pa\g\right)
  \nn
 {\cal K}_n&=&n\g^{n-1}V_{-\al_2}^{\al_1}\left(\g J_\theta-J_{\al_2}\right)-
  \hf\g^n(H_1+2H_2)\nn
 &+&\hf n\g^{n-1}\left(\g^2E_1+\g H_1-F_1-\pa\g\right)
\label{N2gen}
\eea
where $\theta=\al_1+\al_2$.
Note that the contribution $-\hf(k+1/2)n\g^{n-1}\pa\g$ to ${\cal K}_n$
vanishes upon integration as $n\oint\dz\g^{n-1}\pa\g=np\delta_{n,0}=0$.
The $N=2$ SCA becomes
\bea
 \left[L_n,L_m\right]&=&(n-m)L_{n+m}+\frac{c}{12}(n^3-n)\delta_{n+m,0}\nn
 \left[L_n,A_m\right]&=&((h(A)-1)n-m)A_{n+m},\ \ A\in{\{}G,\Gb,K{\}}\nn
 {\{}G_{n+1/2},G_{m+1/2}{\}}&=&{\{}\Gb_{n-1/2},\Gb_{m-1/2}{\}}=0\nn
 {\{}G_{n+1/2},\Gb_{m-1/2}{\}}&=&L_{n+m}+(n-m+1)K_{n+m}+\frac{1}{6}cn(n+1)
  \delta_{n+m,0}\nn
 \left[K_n,G_{m+1/2}\right]&=&\hf G_{n+m+1/2},\ \ \ 
 \left[K_n,\Gb_{m-1/2}\right]=-\hf\Gb_{n+m-1/2}\nn
 \left[K_n,K_m\right]&=&\frac{1}{12}cn\delta_{n+m,0}
\eea
and closure is seen to be ensured by the 4 generators (\ref{N2gen}).
This result has already been obtained by Ito \cite{Ito1}, though his 
construction is based on a slightly different but equivalent free field
realization of the associated affine $SL(2|1)$ current superalgebra.

\section{BRST Invariance}

Before turning to the classical SCA let us discuss an important property
of our construction.
As pointed out in Ref. \cite{GKS}, BRST invariance of the construction
of the space-time conformal algebra from a world sheet $SL(2)$ current algebra
requires the Virasoro generators to be primary fields of weight one
with respect to the world sheet energy-momentum tensor. This ensures that
the integrated fields commute with the world sheet Virasoro algebra.
This requirement carries over to the superconformal case, where 
all (super-)currents (the Virasoro generators ${\cal L}$,
the supercurrents ${\cal G}$ and $\Gbc$, the affine Lie algebra 
generators ${\cal K}$ etc) are primary of weight one with respect to the 
Sugawara energy-momentum tensor of 
the world sheet $SL(2|N/2)$ current superalgebra. A naive inspection 
immediately tells that the four types of currents considered so far
have weight one, so all we
need to verify is that they are primary. This amounts to verifying
that third and higher order poles in the OPEs with the Sugawara
tensor $T$ all vanish. From the free field realization of $T$
it follows that no higher order poles than third order appears.
Using that the affine currents $J$ are primary fields, the BRST
invariance of the supercurrents $G$ is readily confirmed as 
$V_{\al^\pm}^{\al_1}=0$. The BRST invariance of the Virasoro generators
$L$ follows from (\ref{V1}), whereas the invariance of the supercurrents
$\Gb$ is ensured provided
\bea
 0&=&n(n-1)\g^{n-1}\Vm+n(n-1)\g^{n-2}\Vp\nn
 &-&n(n-1)\Vm\left(n\g^{n-1}V_{\al_1}^{\al_1}+(n-1)\g^{n-2}V_1^{\al_1}
  -(n-2)\g^{n-3}V_{-\al_1}^{\al_1}\right)\nn
 &-&n(n-1)\left(\g^nV_{\al_1}^\nu+\g^{n-1}V_1^\nu-\g^{n-2}V_{-\al_1}^\nu
  \right)\pa_\nu\Vm
\eea
The three lines vanish separately due to (\ref{VpVm}), (\ref{V1})
and (\ref{VVm}), respectively.
Likewise, BRST invariance of the affine Lie algebra generators $K$
amounts to verifying
\bea
 0&=&-n\g^{n-1}{f_{\al^-,-\beta^-}}^cV_c^{\al_1}\nn
 &+&\hf\delta_{\al^-,\beta^-}\left\{n\left((n+1)\g^nV_{\al_1}^{\al_1}
  +n\g^{n-1}V_1^{\al_1}-(n-1)\g^{n-2}V_{-\al_1}^{\al_1}\right)
  -n\g^{n-1}V_1^{\al_1}\right\}
\label{TK}
\eea
Here we have used that $V_{\al^\pm}^{\al_1}=0$, and the identity 
(\ref{TK}) follows from (\ref{V1}) and (\ref{fVc}).

One may take a more general point of view observing that the
(anti-)commutator of two BRST invariant fields commutes with the
world sheet Virasoro generators. This follows from the Jacobi
identities. Thus, having established that all but one field
appearing on the right hand side of a (anti-)commutator (of
two BRST invariant fields) are BRST invariant, is sufficient to
conclude that the final field is likewise BRST invariant.
A trivial example is the alternative deduction that $K$ is BRST invariant
following from the anti-commutator (\ref{GGb}).

\section{Classical Superconformal Algebra}

Here we shall distinguish between classical and quantum SCAs.
Our use of the notion {\em quantum} is not in the
quantum group sense of $q$-deformations but rather as opposed to
{\em classical} as described in the following.
Let us recall the situation for free field realizations of affine current 
superalgebras discussed in
Section 3. In that case one may start with a first order linear
differential operator realization of the underlying Lie
superalgebra. The free field realization of the associated current
superalgebra is then obtained by substituting 
with (normal ordered products of) free fields (\ref{subst})
and subsequently
adding ``quantum corrections'', ``anomalous terms'' or ``normal
ordering terms'' (\ref{subst2}). We shall denote
the differential operator realization a classical limit
or version of the associated ``quantum'' free field realization.
A classical algebra thus defined has vanishing central
extensions\footnote{It should be stressed that a non-vanishing central
charge of a classical Virasoro algebra may well exist when classical is
defined to denote single contractions only, as $\pa^2\var(z)\pa^2\var(w)
=-12/(z-w)^4$. Here we have used the convention $\var(z)\var(w)=2\ln(z-w)$.
However, terms like $\pa^2\var$ are excluded in our
``differential operator realization picture'' employed in the
present paper. Note that BRST invariance is used as an implicit
guideline as $\pa^2\var$ has weight 2 with respect to the world sheet
energy-momentum tensor.}.

A similar situation may be expected in the present case.
Thus, there should exist a classical counterpart of the full SCA
which allows a differential operator realization (and accordingly
has vanishing
central extensions). Based on this assumption our program is to
first work out the classical SCA for then to perform the appropriate
substitutions and additions of anomalous terms in order to obtain
the full (quantum) SCA. It should be stressed that to each {\em mode} of the 
generators of the quantum SCA, there is an associated differential operator.
This results in an infinite dimensional algebra of differential
operators contrary to the situation described above where the classical
algebra is a standard (finite dimensional) Lie superalgebra.

Having identified the differential operator
\ben
 A(x,\theta,\pa,\La)=\sum_a Y^a(x,\theta)J_a(x,\theta,\pa,\La)
\label{Aansatz}
\een
as a generator of the classical SCA, the corresponding quantum generator
\ben
 A=\oint\dz{\cal A}(z)
\label{A}
\een
is obtained by performing the substitutions (\ref{subst}) in
$A(x,\theta,\pa,\La)$ and adding appropriate anomalous terms linear in 
derivatives of the spin 0 ghost fields 
in order to produce ${\cal A}(z)$:
\bea
 {\cal A}(z)&=&\sum_a Y^a(\g(z),c(z))J_a(z)\nn
  &+&\sum_{\al\in\D^0_+}X_\al(\g(z),c(z))\pa\g^\al(z)
  +\sum_{\al\in\D^1_+}X_\al'(\g(z),c(z))\pa c^\al(z)
\label{Aext}
\eea
So with the ansatz (\ref{Aansatz}), ${\cal A}(z)$ is linear in the 
affine currents $J_a(z)$ with spin 0 ghost field dependent coefficients.
Note that in the expression (\ref{Aext}) some anomalous terms
are ``hidden'' in the definition of $J_a(z)$, cf. (\ref{subst2}). 

The question of BRST invariance of $A$ (\ref{A})
may be addressed even without explicit
knowledge on the anomalous term. This follows from the fact
that any term of the form $\sum_\al Z_\al(\g,c)\pa\g^\al+\sum_\al Z_\al'(\g,c)
\pa c^\al$ is primary of weight one.
Thus, in order to establish that $A$ is BRST invariant it suffices to
consider the term linear in the affine (super-)currents.
In the following we shall accordingly define a classical differential
operator to be BRST invariant when its ``naively quantized'' form linear
in the affine (super-)currents is BRST invariant.

Before continuing our program let us briefly justify it. It has turned
out to be an immense technical task to complete the derivation
of the SCA for general $N$. Even at the classical level, the computations are 
rather involved. A study of the center-less classical SCA seems therefore
a natural first project to concentrate on,
and one from which one may get structural
insight into the full quantum SCA. In the following we shall
present some essential steps in the direction
of deriving the classical SCA. In the presumably most interesting
case of $N=4$, the classical SCA is completed. The full quantum
$N=4$ with generic central charge will be presented elsewhere
\cite{Ras2}.

\subsection{Algebra Generators}

In the remaining part of this Section all fields $A$ are represented by
their classical differential operator analogues $A(x,\theta,\pa,\La)$.
To be explicit, let us summarize our findings for the classical
generators:
\bea
 L_n&=&a_+(n)x^{n+1}E_1+a_3(n)x^nH_1+a_-(n)x^{n-1}F_1\nn
 G_{\al^-;n+1/2}&=&(n+1)x^nJ_{\al^-}-nx^{n+1}J_{\al^+}\nn
 \Gb_{-\al^-;n-1/2}&=&(n-1)x^nJ_{-\al^-}+nx^{n-1}J_{-\al^+}\nn
 &-&n(n-1)x^{n-2}\Vm\left(x^2E_1+xH_1-F_1\right)\nn
 K_{\al^-;-\beta^-;n}&=&nx^{n-1}\Vmb\left(xJ_{\al^+}-J_{\al^-}\right)
  -x^n{f_{\al^-,-\beta^-}}^cJ_c\nn
 &+&\hf\delta_{\al^-,\beta^-}x^{n-1}\left(nx^2E_1+(n-1)xH_1-nF_1\right)\nn
 h(L,G,\Gb,K)&=&(2,3/2,3/2,1)
\label{LGGbKc}
\eea
$J_a$ ($E$, $H$ or $F$) denotes the differential operator 
$J_a(x,\theta,\pa,\La)$ given in (\ref{diff}), 
(\ref{pol}) while $V_{-\beta^-}^{\al_1}$ is the polynomial in the
super-triangular coordinates $x$ and $\theta$ given in (\ref{pol}). 
Here and in the following $x$ may denote either $x^{\al_1}$ or
a general triangular coordinate argument, though 
it should be clear from the context
which it is. $h(A_m)$ indicates that $A_m$ is primary of weight $h$:
\ben
 \left[L_n,A_m\right]=((h(A)-1)n-m)A_{n+m}
\een
The generators respect among others the anti-commutators
\bea
 \left\{G_{\al^-;n+1/2},G_{\beta^-;m+1/2}\right\}&=&0\nn
 \left\{G_{\al^-;n+1/2},\Gb_{-\beta^-;m-1/2}\right\}
 &=&\delta_{\al^-,\beta^-}L_{n+m}+(n-m+1)K_{\al^-;-\beta^-;n+m}
\label{LGGbKc2}
\eea 

We shall now discuss the subalgebra generated 
by $\left\{K\right\}$. One finds straightforwardly
\ben
 \left[K_{\al^-;-\beta^-;n},K_{\mu^-;-\nu^-;m}\right]=\delta_{\mu^-,\beta^-}
  K_{\al^-;-\nu^-;n+m}-\delta_{\al^-,\nu^-}K_{\mu^-;-\beta^-;n+m}
\label{KKc}
\een
In order to show explicitly that this has the affine structure
\ben
 SL(N/2)\otimes U(1)
\label{SLU}
\een
we introduce the following notation. From Appendix A we know that
any root $\al^-\in\D_+^{1-}$ may be represented as $\eps_2-\delta_u$
for some $u=1,...,N/2$, so abbreviate $K$ by
\ben
 K_{u;v;n}=K_{\eps_2-\delta_u;-(\eps_2-\delta_v);n}
\een
Note also that $\delta_v-\delta_u>0$ for $u>v$. Define now
\bea
 \tilde{E}_{i;n}&=&K_{i+1;i;n}\nn
 \tilde{H}_{i;n}&=&K_{i+1;i+1;n}-K_{i;i;n}\nn
 \tilde{F}_{i;n}&=&K_{i;i+1;n}
\eea
where $i=1,...,N/2-1$ by construction. One may then show that these
correspond to the Chevalley generators of an (center-less)
affine $SL(N/2)$ Lie algebra.
In general, the currents $K_{u;v;n}$ correspond to raising operators
for $u>v$, and to lowering operators for $u<v$.
Furthermore, the generator
\ben
 U_n=\sum_{u=1}^{N/2}K_{u;u;n}
\een
is seen to commute with all ladder operators $K_{u;v\neq u;m}$, with the
Cartan generators $\tilde{H}_{i,m}$ and with $U_m$ itself. 
Thus, $U$ generates a (center-less)
$U(1)$ current algebra and we have the decomposition (\ref{SLU}).

Let us return to the anti-commutator 
$\left\{\Gb_{-\al^-},\Gb_{-\beta^-}\right\}$ and prove
the classical counterpart of the assertion (\ref{GbGb}).
The anti-commutator may be computed directly or obtained as a special
case of a much more general consideration: Introduce the operator
\bea
 \Gb_{-\beta_1^-,...,-\beta_k^-;m-1+k/2}
 &=&V_{-\beta_1^-}^{\al_1}...V_{-\beta_{k-1}^-}^{\al_1}
  \left\{(m-2+k)x^{m}J_{-\beta_k^-}+mx^{m-1}J_{-\beta_k^+}\right\}\nn
 &+&...\nn
 &\vdots&\nn
 &+&\left\{(m-2+k)x^{m}J_{-\beta_1^-}+mx^{m-1}J_{-\beta_1^+}
  \right\}V_{-\beta_2^-}^{\al_1}...V_{-\beta_{k}^-}^{\al_1}\nn
 &-&V_{-\beta_1^-}^{\al_1}...V_{-\beta_{k}^-}^{\al_1}
  \left\{(m+k-1)(m+k-2)x^{m}E_1\right.\nn
 &&\left.+(m+k-2)mx^{m-1}H_1-m(m-1)x^{m-2}F_1\right\}
\label{Gbb}
\eea
which is seen to reduce to $\Gb_{-\beta^-}$ for $k=1$.  
$\Gb_{-\beta_1^-,...,-\beta_k^-;m}$ is bosonic (fermionic) for $k$ 
even (odd). In the latter notation 
the mode $m$ is meant to be integer or half-integer depending on
the parity of the generator, i.e. depending on $k$ being even or odd,
respectively. Note that $\Gb_{-\beta_1^-,...,-\beta_k^-}$ is 
anti-symmetric in its root indices.
$J_{-\beta_j^\pm}$ is defined {\em not} to act on 
$V_{-\beta_{j+1}^-}^{\al_1}...V_{-\beta_{k}^-}^{\al_1}$, and is only
written to the left of the $V$-monomial for convenience of notation.
Thus, within $\Gb_{-\beta_1^-,...,-\beta_k^-}$ one has
$J_{-\beta_j^\pm}V_{-\beta_{j+1}^-}^{\al_1}...V_{-\beta_{k}^-}^{\al_1}
=(-1)^{k-j}V_{-\beta_{j+1}^-}^{\al_1}...V_{-\beta_{k}^-}^{\al_1}
J_{-\beta_j^\pm}$. This resembles normal ordering needed in the free
field realization and may be regarded as a {\em normal ordering of 
the differential operator}. It is to be employed throughout this section. 
One may now work out the (anti-)commutator
\ben
 \left[\Gb_{-\beta_1^-,...,-\beta_k^-;n},
  \Gb_{-\la_1^-,...,-\la_l^-;m}\right\}
 =\left((k-2)m-(l-2)n\right)
  \Gb_{-\beta_1^-,...,-\beta_k^-,-\la_1^-,...,-\la_l^-;n+m}
\label{GbGl}
\een
We observe that 
$\left\{\Gb_{-\beta_1^-,...,-\beta_k^-}\right\}_{k=1}^{|\D_-^{1-}|}$
generate a subalgebra of dimension $2^{|\D_-^{1-}|}-1=2^{N/2}-1$ and that
$2^{N/2-1}$ of the generators are fermionic. Note that the commutator for
$k=l=2$ vanishes identically.
One may also show that (classically) the generators are primary:
\ben
 \left[L_n,\Gb_{-\beta_1^-,...,-\beta_k^-;m}\right]=((1-k/2)n-m)
  \Gb_{-\beta_1^-,...,-\beta_k^-;n+m}\ , \ \ \ \ h=2-k/2
\label{LGbb}
\een
Thus, including $L$ as a generator of the subalgebra it has dimension
$2^{|\D_-^{1-}|}$ and equal numbers of bosonic and fermionic generators.
BRST invariance is readily verified, either directly or as a consequence 
of the recursive relation (\ref{GbGl}) and the general approach of Section 5.

We note that the generator (\ref{Gbb}) may be written in the
following compact form
\bea
 \Gb_{-\beta_1^-,...,-\beta_k^-;m-1+k/2}
 &=&x^{m-2}{\{}\sum_{j=1}^k\left(
  (m+k-2)x^2J_{-\beta_j^-}+mxJ_{-\beta_j^+}\right)\frac{\pa}{\pa 
  V_{-\beta_j^-}^{\al_1}}\nn
 &&-(m+k-1)(m+k-2)x^2E_1-(m+k-2)mxH_1\nn
 &&+m(m-1)F_1{\}}
 \left(V_{-\beta_1^-}^{\al_1}...V_{-\beta_k^-}^{\al_1}\right)
\eea
where we have defined
\ben
 \frac{\pa}{\pa V_{-\beta_j^-}^{\al_1}}V_{-\beta_i^-}^{\al_1}\equiv\delta_{ij}
\label{ddVV}
\een
As $\Gb_{-\beta_1^-,...,-\beta_k^-}$ is a first order
differential operator, $\pa/\pa V_{-\beta_j^-}^{\al_1}$ is meant
to act only on the explicitly written products of $V$'s within 
$\Gb_{-\beta_1^-,...,-\beta_k^-}$ itself.
A special situation occurs for $k=2$ since we then have
\bea
 \Gb_{-\beta_1^-,-\beta_2^-;m}&=&mS_{-\beta_1^-,-\beta_2^-;m}\nn
 S_{-\beta_1^-,-\beta_2^-;m}&=&x^{m-2}{\{}\sum_{j=1}^2\left(x^2J_{-\beta_j^-}
  +xJ_{-\beta_j^+}\right)\frac{\pa}{\pa V_{-\beta_j^-}^{\al_1}}\nn
 &&-(m+1)x^2E_1-mxH_1+(m-1)F_1{\}}\left(V_{-\beta_1^-}^{\al_1}
  V_{-\beta_2^-}^{\al_1}\right)
\label{k2scalar}
\eea
and $S_{-\beta_1^-,-\beta_2^-}$ is readily seen to have weight 0.
$\Gb_{-\beta_1^-,-\beta_2^-}$ may be interpreted as the derivative of the 
scalar $S_{-\beta_1^-,-\beta_2^-}$. 

The list of generators presented hitherto is by no means exhaustive.
Let us consider generators which may
be obtained by the adjoint action of $\left\{G_{\al^-}\right\}$ on
$\Gb_{-\beta_1^-,...,-\beta_k^-}$. 
Firstly, we find the (anti-)commutator
\bea
 \left[G_{\al^-;n+1/2},\Gb_{-\beta_1^-,...,-\beta_k^-;m}\right\}&=&
  (-m-(k-2)n)\Phi_{\al^-;-\beta_1^-,...,-\beta_k^-;n+m+1/2}\nn
 &+&\frac{k-2}{k-3}\sum_{j=1}^k(-1)^{j-1}\delta_{\al^-,\beta_j^-}
  \Gb_{-\beta_1^-,...,\widehat{-\beta_j^-},...,-\beta_k^-;n+m+1/2}
\eea
where $\Phi_{\al^-;-\beta_1^-,...,-\beta_k^-}$ corresponds
to the special case $l=1$ in the following general expression
\bea
 &&\Phi_{\mu_1^-,...,\mu_l^-;-\beta_1^-,...,-\beta_k^-;m}\nn
 &=&x^{m-(k-l)/2-1}\sum_{i=1}^l(-1)^{i-1}{\{}(m-(k-l)/2)J_{\mu_i^-}
  -(m+(k-l)/2)xJ_{\mu_i^+}\nn
 &&-x\sum_{j=1}^k{f_{\mu_i^-,-\beta_j^-}}^cJ_c
  \frac{\pa}{\pa V_{-\beta_j^-}^{\al_1}}{\}}
  \frac{\pa}{\pa V_{-\mu_1^-}^{\al_1}}...
  \widehat{\frac{\pa}{\pa V_{-\mu_i^-}^{\al_1}}}...
  \frac{\pa}{\pa V_{-\mu_l^-}^{\al_1}}\left(
   V_{-\beta_1^-}^{\al_1}...V_{-\beta_{k}^-}^{\al_1}\right)\nn
 &+&\frac{l}{k-l-2}x^{m-(k-l)/2-1}
  {\{}\sum_{j=1}^k\left(x^2J_{-\beta_j^-}
  +xJ_{-\beta_j^+}\right)\frac{\pa}{\pa V_{-\beta_j^-}^{\al_1}}\nn
 &&-(m+(k-l)/2)x^2E_1-(m+(k-l)/2-1)xH_1\nn
 &&+(m-(k-l)/2)F_1
  {\}}\frac{\pa}{\pa V_{-\mu_1^-}^{\al_1}}...
  \frac{\pa}{\pa V_{-\mu_l^-}^{\al_1}}\left(
   V_{-\beta_1^-}^{\al_1}...V_{-\beta_{k}^-}^{\al_1}\right)
\label{Phi}
\eea
These generators are only defined for certain integer pairs $(l,k)$ 
to be discussed
below. A hat over an object indicates that the object is left out.
We observe that 
$\Phi_{\mu_1^-,...,\mu_l^-;-\beta_1^-,...,-\beta_k^-}$ is
bosonic (fermionic) for $l+k$ even (odd), and that it is anti-symmetric
in the positive root indices and in the negative root indices, separately. 
Using the polynomial relations
listed at the end of Appendix A, one may show that 
$\Phi$ satisfies the recursion relation
\bea
 &&\left[G_{\nu^-;n+1/2},\Phi_{\mu_1^-,...,\mu_l^-;-\beta_1^-,...,-\beta_k^-
  ;m}\right\}\nn
 &=&\frac{k-2}{k-l-2}\sum_{j=1}^k(-1)^{l-j+1}\delta_{\nu^-,\beta_j^-}
  \Phi_{\mu_1^-,...,\mu_l^-;-\beta_1^-,...,\widehat{-\beta_j^-},...,
  -\beta_k^-;n+m+1/2}\nn
 &-&\frac{l}{k-l-2}\Phi_{\nu^-,\mu_1^-,...,\mu_l^-;-\beta_1^-,...,-\beta_k^-;
  n+m+1/2}
\label{GPhi}
\eea
In addition, $\Phi_{\mu_1^-,...,\mu_l^-;-\beta_1^-,...,-\beta_k^-}$ 
may be shown to be primary of weight 
\ben
 h(\Phi_{\mu_1^-,...,\mu_l^-;-\beta_1^-,...,-\beta_k^-})=1-(k-l)/2
\een
As for $\Gb_{-\beta_1^-,...,-\beta_k^-}$ (\ref{Gbb}),
BRST invariance of $\Phi_{\mu_1^-,...,\mu_l^-;-\beta_1^-,...,-\beta_k^-}$ 
is verified straightforwardly, either directly or indirectly.
There is a slight subtlety in (\ref{GPhi}) for $k=l=1$, which is easily
resolved, though, as $\Phi_{\mu^-}$ (\ref{Phi})
may be interpreted as $G_{\mu^-}$. One then has
\ben
 \left[K_{\mu^-;-\beta^-;n},G_{\nu^-;m+1/2}\right]=\delta_{\nu^-,\beta^-}
  G_{\mu^-;n+m+1/2}-\hf\delta_{\mu^-,\beta^-}G_{\nu^-;n+m+1/2}
\een

In the definition of $\Phi_{\mu_1^-,...,\mu_l^-;-\beta_1^-,...,-\beta_k^-}$ 
we obviously have $k-l-2\neq0$, and as the expression 
is obtained by the adjoint action of $G$ on
$\Gb_{-\beta_1^-,...,-\beta_k^-}$, $1\leq l<k-2$ is seen to be
a valid domain. It is relevant for the discussion below on $N=4$ that
besides $(l,k)=(1,1)$ which corresponds to the $K$ generators, also
$(l,k)\in{\{}(2,1),(1,2),(2,2),(3,2){\}}$ 
may be reached by this adjoint action. 
It turns out that three of these four generators may be expressed
in terms of simpler generators. First we note that up to permutations
in the root indices, 
$\Phi_{\mu^-,\beta_2^-,\beta_1^-;-\beta_1^-,-\beta_2^-}$ is the only
non-vanishing generator of type $(3,2)$ (provided $\mu^-\neq\beta_i^-$
and $\beta^-_1\neq\beta_2^-$, of course), while
$\Phi_{\mu^-,\beta^-;-\beta^-}$ is the only non-vanishing generator
of type $(2,1)$ (provided $\mu^-\neq\beta^-$). 
It is easily shown that they satisfy
\bea
 \Phi_{\mu^-_1,\mu_2^-,\mu_3^-;-\beta_1^-,-\beta_2^-;m+1/2}&=&\left(
  \delta_{\mu^-_3,\beta_1^-}\delta_{\mu^-_2,\beta_2^-}
  -\delta_{\mu^-_3,\beta_2^-}\delta_{\mu^-_2,\beta_1^-}\right)
  G_{\mu_1^-;m+1/2}\nn
 &-&\left(\delta_{\mu^-_3,\beta_1^-}\delta_{\mu^-_1,\beta_2^-}
  -\delta_{\mu^-_3,\beta_2^-}\delta_{\mu^-_1,\beta_1^-}\right)
  G_{\mu_2^-;m+1/2}\nn
 &+&\left(\delta_{\mu^-_2,\beta_1^-}\delta_{\mu^-_1,\beta_2^-}
  -\delta_{\mu^-_2,\beta_2^-}\delta_{\mu^-_1,\beta_1^-}\right)
  G_{\mu_3^-;m+1/2}\nn
 \Phi_{\mu^-_1,\mu_2^-;-\beta^-;m+1/2}&=&\delta_{\mu_2^-,\beta^-}G_{\mu_1^-}
  -\delta_{\mu_1^-,\beta^-}G_{\mu_2^-}
\label{l3k2}
\eea
Secondly, we observe that $\Phi_{\mu^-,\al^-;-\al^-,-\beta^-}$
is the only non-vanishing generator of type $(2,2)$ and that it
may be reduced as
\bea
 \Phi_{\mu^-_1,\mu_2^-;-\beta_1^-,-\beta_2^-;m}&=&
  -\delta_{\mu_1^-,\beta_1^-}K_{\mu_2;-\beta_2^-;m}
  +\delta_{\mu_1^-,\beta_2^-}K_{\mu_2;-\beta_1^-;m}\nn
 &+&\delta_{\mu_2^-,\beta_1^-}K_{\mu_1;-\beta_2^-;m}
  -\delta_{\mu_2^-,\beta_2^-}K_{\mu_1;-\beta_1^-;m}
\label{l2k2}
\eea
For $l>k+1$, $k=1,2$, the generator 
$\Phi_{\mu_1^-,...,\mu_l^-;-\beta_1^-,...,-\beta_k^-}$ is readily seen
to vanish.
Relevant for the discussion on the $N=4$ SCA in the following, is the
result
\bea
 \left[K_{\mu^-;-\nu^-;n},\Gb_{-\al^-;m-1/2}\right]&=&\hf\delta_{\mu^-,\nu^-}
  \Gb_{-\al^-;n+m-1/2}-\delta_{\mu^-,\al^-}\Gb_{-\nu^-;n+m-1/2}\nn
 &+&n\Phi_{\mu^-;-\nu^-,-\al^-;n+m-1/2}
\eea
which one may work out explicitly.

Finally, let us add a comment on the closure of the algebra. Based on the
results obtained so far one should expect that the SCA is generated
by a BRST invariant set ${\{}\tilde{\Phi}_{\mu_1^-,...,\mu_l^-;-\beta_1^-,
...,-\beta_k^-}{\}}$ where $0\leq l,k$ and $l\leq k+1$. The notation implies 
for example $\tilde{\Phi}_\mu=G_\mu$ and $\tilde{\Phi}_{-\beta_1^-,
...,-\beta_k^-}=\Gb_{-\beta_1^-,...,-\beta_k^-}$, while for $l=k=0$
we have $\tilde{\Phi}=L$. Not all the fields $\tilde{\Phi}$ need be present
as independent fields
as illustrated by the reductions (\ref{l3k2}) and (\ref{l2k2}).
At present we do not have a complete proof of the assumption that 
${\{}\tilde{\Phi}{\}}$ generates the SCA, though it is easy to prove
(using the polynomial relations of Appendix A)
that the following 8 BRST non-invariant ``building blocks''
\bea
 &V_{-\beta_1^-}^{\al_1}...V_{-\beta_{k_+^-}^-}^{\al_1}x^{n_+^-}J_{\al^-},\ \ 
  V_{-\beta_1^-}^{\al_1}...V_{-\beta_{k_+^+}^-}^{\al_1}x^{n_+^+}J_{\al^+},\ \ 
  V_{-\beta_1^-}^{\al_1}...\widehat{V_{-\beta_j^-}^{\al_1}}...
  V_{-\beta_{k_c}^-}^{\al_1}x^{n_c}{f_{\al^-,-\beta_j^-}}^cJ_c&\nn
 &V_{-\beta_1^-}^{\al_1}...V_{-\beta_{k_-^-}^-}^{\al_1}x^{n_-^-}
  J_{-\al^-},\ \
  V_{-\beta_1^-}^{\al_1}...V_{-\beta_{k_-^+}^-}^{\al_1}x^{n_-^+}
  J_{-\al^+}&\nn
 &V_{-\beta_1^-}^{\al_1}...V_{-\beta_{k_E}^-}^{\al_1}x^{n_E}E_1,\ \ 
  V_{-\beta_1^-}^{\al_1}...V_{-\beta_{k_H}^-}^{\al_1}x^{n_H}H_1,\ \ 
  V_{-\beta_1^-}^{\al_1}...V_{-\beta_{k_F}^-}^{\al_1}x^{n_F}F_1&
\eea
close under (anti-)commutations. 

Despite the fact that closure of the BRST invariant
algebra is still not ensured, nor are all the BRST invariant
generators at hand, above we have presented substantial
evidence that our construction does produce a finitely generated and BRST
invariant $N$ extended SCA. Below we shall demonstrate that this
is indeed the case for $N=4$.
We intend to come bask elsewhere with further discussion on the BRST
invariant SCA and its quantization as described above.

As $|\D_-^{1-}|=N/2$, we observe that for $N=6,8,...$ the SCA contains
primary generators of negative weight causing problems for the unitarity
of the associated CFT, and in particular for its applications to string
theory. From an algebraic point of view, however, we see no severe
obstacles arising from the appearance of negative weights, and believe
that further studies of these algebras are warranted.
Their explicit realizations and linearity in the affine currents
seem to make such investigations feasible.

\subsection{Classical $N=4$ Superconformal Algebra}

We recall that $sl(2|2)$ has precisely two positive
(fermionic) $\al^-$-roots, see Appendix A:
\ben
 \D_+^{1-}={\{}\al_2,\al_{2+3}\equiv\al_2+\al_3{\}}\ 
  ,\ \ \ \ \ \al_2=\eps_2-\delta_1\ ,\ \ \al_3=\delta_1-\delta_2
\een
Using the results on (classical) SCA obtained above for general $N$
we may almost immediately complete the (classical)
$N=4$ SCA. We find that closure
is ensured by the following 12 BRST invariant generators which may
be characterized as: 
\bea
 \mbox{Virasoro\ generator}\ \ &L&\ \ h=2\nn
 \mbox{supercurrents}\ \ &G_{\al_2},\ G_{\al_{2+3}},\ \Gb_{-\al_2},\ 
  \Gb_{-\al_{2+3}}&\ \ h=3/2\nn
 \mbox{affine}\ SL(2)\ \ &\tilde{E}=K_{\al_{2+3};-\al_2},&\nn 
 &\tilde{H}=K_{\al_{2+3};-\al_{2+3}}-K_{\al_{2};-\al_2},&\nn
  &\tilde{F}=K_{\al_{2};-\al_{2+3}}&\ \ h=1\nn
 \mbox{affine} \ U(1)\ \ &U=K_{\al_{2};-\al_2}+K_{\al_{2+3};-\al_{2+3}}&\ \ 
  h=1\nn
 \mbox{fermions}\ \ &\phi_{-\al_2}=\Phi_{\al_{2+3};-\al_{2+3},-\al_2},&\nn 
 &\phi_{-\al_{2+3}}=\Phi_{\al_{2};-\al_{2},-\al_{2+3}}&\ \ h=1/2\nn
 \mbox{scalar}\ \ &S&\ \ h=0\nn
 &(nS_n=\Gb_{-\al_2,-\al_{2+3};n}&\ \ h=1)
\label{N4gen}
\eea
The non-trivial (anti-)commutators are
\bea
 \left[L_n,A_m\right]&=&((h(A)-1)n-m)A_{n+m}\nn
 \left\{G_{\al^-;n+1/2},G_{\beta^-;m+1/2}\right\}&=&0\nn
  \left\{G_{\al^-;n+1/2},\Gb_{-\beta^-;m-1/2}\right\}&=&\delta_{\al^-,\beta^-}
   L_{n+m}+(n-m+1)K_{\al^-;-\beta^-;n+m}\nn
 \left\{\Gb_{-\al_2;n-1/2},\Gb_{-\al_{2+3};m-1/2}\right\}&=&(n-m)
  (n+m-1)S_{n+m-1}\nn
 \left[\tilde{H}_n,\tilde{E}_m\right]&=&2\tilde{E}_{n+m},\ \ 
 \left[\tilde{H}_n,\tilde{F}_m\right]=-2\tilde{F}_{n+m},\ \ 
 \left[\tilde{E}_n,\tilde{F}_m\right]=\tilde{H}_{n+m}\nn
 \left[\tilde{E}_n,G_{\al_2;m+1/2}\right]&=&G_{\al_{2+3};n+m+1/2},\ \ 
 \left[\tilde{F}_n,G_{\al_{2+3};m+1/2}\right]=G_{\al_2;n+m+1/2}\nn
 \left[\tilde{H}_n,G_{\al_2;m+1/2}\right]&=&-G_{\al_2;n+m+1/2},\ \
 \left[\tilde{H}_n,G_{\al_{2+3};m+1/2}\right]=G_{\al_{2+3};n+m+1/2}\nn
 \left[\tilde{E}_n,\Gb_{-\al_{2+3};m-1/2}\right]&=&
  -\Gb_{-\al_{2};n+m-1/2}-n\phi_{-\al_2;n+m-1/2}\nn
 \left[\tilde{H}_n,\Gb_{-\al_{2};m-1/2}\right]&=&
  \Gb_{-\al_{2};n+m-1/2}+n\phi_{-\al_2;n+m-1/2}\nn
 \left[\tilde{H}_n,\Gb_{-\al_{2+3};m-1/2}\right]&=&
  -\Gb_{-\al_{2+3};n+m-1/2}-n\phi_{-\al_{2+3};n+m-1/2}\nn
 \left[\tilde{F}_n,\Gb_{-\al_{2};m-1/2}\right]&=&
  -\Gb_{-\al_{2+3};n+m-1/2}-n\phi_{-\al_{2+3};n+m-1/2}\nn
 \left[U_n,\Gb_{-\al_{2};m-1/2}\right]&=&n\phi_{-\al_2;n+m-1/2}\nn
 \left[U_n,\Gb_{-\al_{2+3};m-1/2}\right]&=&
  n\phi_{-\al_{2+3};n+m-1/2}\nn
 \left[S_{n},G_{\al_2;m+1/2}\right]&=&\phi_{-\al_{2+3};n+m+1/2},\ \ 
 \left[S_{n},G_{\al_{2+3};m+1/2}\right]=-\phi_{-\al_{2};n+m+1/2}\nn 
 \left\{G_{\al_2;n+1/2},\phi_{-\al_2;m-1/2}\right\}&=&U_{n+m},\ \
 \left\{G_{\al_{2+3};n+1/2},\phi_{-\al_{2+3};m-1/2}\right\}=U_{n+m}\nn
 \left\{\Gb_{-\al_2;n-1/2},\phi_{-\al_{2+3};m-1/2}\right\}&=&
  (n+m-1)S_{n+m-1}\nn
 \left\{\Gb_{-\al_{2+3};n-1/2},\phi_{-\al_{2};m-1/2}\right\}&=&
  -(n+m-1)S_{n+m-1}\nn 
 \left[\tilde{E}_n,\phi_{-\al_{2+3};m-1/2}\right]&=&-\phi_{-\al_{2};n+m-1/2}
  ,\ \ 
 \left[\tilde{F}_n,\phi_{-\al_{2};m-1/2}\right]=-\phi_{-\al_{2+3};n+m-1/2}\nn
 \left[\tilde{H}_n,\phi_{-\al_2;m-1/2}\right]&=&\phi_{-\al_2;n+m-1/2},\ \
 \left[\tilde{H}_n,\phi_{-\al_{2+3};m-1/2}\right]=-\phi_{-\al_{2+3};n+m-1/2}
\nn
\label{N4}
\eea
$A_m$ denotes any of the 12 BRST invariant generators listed in
(\ref{N4gen}).
We observe that only the derivative of the scalar $S$ appears
on the right hand sides of (\ref{N4}). 
Thus, the zero mode of $S$ decouples from the algebra. 
One may verify explicitly that the Jacobi identities are satisfied.

Finally, we note that this center-less
$N=4$ SCA is of a new and asymmetric form. 
In particular, it deviates essentially from the small $N=4$ SCA announced
in Ref. \cite{Ito1} to be obtained by a similar construction. Even though
the free field realization of the associated affine $SL(2|2)$ current
superalgebra used in Ref. \cite{Ito1} is slightly different from ours,
one may show that the result in Ref.
\cite{Ito1} for the $N=4$ SCA is incorrect.
One way of reaching this conclusion is to consider the analogue to
our (\ref{GbGb}) and specialize to the case $n=0$. In the notation of Ref.
\cite{Ito1}, this corresponds to considering the anti-commutator
${\{}\Gb_{-1/2}^1,\Gb_{m-1/2}^2{\}}$ in which case $\Gb_{-1/2}^1$
reduces to $\oint\dz j_{\al_1+\al_2}$ 
(still in the notation of Ref. \cite{Ito1}).
We find that the anti-commutator for generic $m$ is {\em non-vanishing} 
in agreement
with our result but contrary to the definition of the small $N=4$ SCA. 
Nevertheless, the small $N=4$ SCA {\em can} be obtained by a construction
equivalent to the one employed above. One simply has to replace the
original world sheet $SL(2|2)$ current superalgebra by an $SL(2|2)/U(1)$
current superalgebra, whereby the resulting space-time SCA reduces to
the standard small $N=4$ SCA. This will be discussed 
further in Ref. \cite{Ras2}.

\section{Conclusion}

In the present paper a new class of two-dimensional $N$ extended
SCAs has been discussed. The algebras are induced
by free field realizations of affine $SL(2|N/2)$ current
superalgebras, where $N$ is even. In the framework of string theory
on $AdS_3$ the affine $SL(2|N/2)$ current
superalgebra resides on the world sheet providing a space-time SCA 
on the boundary of $AdS_3$. The construction generalizes 
recent work by Ito \cite{Ito1}.
The Virasoro generators, the $N$ supercurrents, and the generators
of an internal
$SL(N/2)\otimes U(1)$ Kac-Moody algebra have all been constructed
explicitly. Reducing the considerations to a classical center-less
limit has provided additional insight into the structure of the full
SCA. BRST invariance has also been addressed.
The classical $N=4$ SCA is complete and of a new type. In particular,
it differs from the small $N=4$ SCA. It also illustrates the new and
important property of the general construction that it treats 
the supercurrents asymmetrically.

The results presented here
offer (``stringy'') representations of superconformal
algebras which are {\em linear} in the currents. This suggests that they
may be useful when discussing representation theoretical questions,
and in the computation of correlation functions. Many other applications
may be envisaged. 

Several classes of Lie supergroups enjoy decompositions of the bosonic 
part $G=SL(2)\otimes G'$ as in the case of $SL(2|N/2)$. Based on their
associated current superalgebras, we anticipate that other classes
of SCAs may be constructed along the lines
employed in the present paper. This is currently being investigated.

In the classification of CFT with extended symmetries, the construction
of SCAs in the present paper presents an alternative to conventional
Hamiltonian reduction and otherwise constructed non-linearly extended
SCAs \cite{Kni,Ber,Sch2,Bow,IMP,FL,STT}. 
Whole new classes of extended Virasoro algebras
seem to be the result of it. There are strong indications that we are
even able to produce new and purely bosonic (and linear) extensions
of the Virasoro algebra. These will be the subject of a forthcoming
publication.
\\[.4cm]
{\bf Acknowledgment}\\[.2cm]
The author is grateful to K. Olsen and J.L. Petersen for comments at
early stages of this work, 
and thanks The Niels Bohr Institute, where parts of this work
were done, for its kind hospitality. 

\appendix
\section{Lie Superalgebra $sl(2|M)$}

The root space of the Lie superalgebra $sl(2|M)$ in the distinguished
representation may be realized in terms
of an orthonormal two-dimensional basis $\left\{\eps_1,\eps_2\right\}$
and an orthonormal $M$-dimensional basis $\left\{\delta_u\right\}_{u=1,...,M}$
with metrics
\ben
 \eps_{\iota}\cdot\eps_{\iota'}=\delta_{\iota,\iota'}
  \spa\delta_u\cdot\delta_{u'}=-\delta_{u,u'}\spa\eps_\iota\cdot\delta_u=0
\een
The $\hf(M+1)(M+2)$ positive roots are then represented as
\bea
 \D^0_+&=&\left\{\eps_1-\eps_2\right\}\cup\left\{\delta_u-\delta_v\ |\ 
  u<v\right\}\nn
 \D^{1+}_+&=&\left\{\eps_1-\delta_u\ |\ u=1,...,M\right\}\nn
 \D^{1-}_+&=&\left\{\eps_2-\delta_u\ |\ u=1,...,M\right\}
\eea
where the $M+1$ simple roots $\al_i$ are 
\bea
 \al_1&=&\eps_1-\eps_2\nn
 \al_2&=&\eps_2-\delta_1\nn
 \al_{u+2}&=&\delta_u-\delta_{u+1}
\eea
The associated ladder operators $E_\al,F_\al$, and the Cartan generators 
$H_i$ admit 
a standard oscillator realization (see e.g. \cite{Tan})
\bea
 E_{\eps_1-\eps_2}=\adg_1a_2\spa &E_{\eps_\iota-\delta_u}=\adg_\iota b_u\spa
  &E_{\delta_u-\delta_v}=\bdg_ub_v\nn
 F_{\eps_1-\eps_2}=\adg_2a_1\spa &F_{\eps_\iota-\delta_u}=\bdg_ua_\iota\spa
  &F_{\delta_u-\delta_v}=\bdg_vb_u\nn
 H_1=\adg_1a_1-\adg_2a_2\spa &H_2=\adg_2a_2+\bdg_1b_1\spa
  &H_{u+2}=\bdg_ub_u-\bdg_{u+1}b_{u+1}
\label{osc}
\eea
where $a^{(\dagger)}_\iota$ and $b^{(\dagger)}_u$ are fermionic and bosonic
oscillators, respectively, satisfying
\ben
 \left\{a_\iota,\adg_{\iota'}\right\}=\delta_{\iota,\iota'}
  \spa \left[b_u,\bdg_v\right]
  =\delta_{u,v}\spa \left[b^{(\dagger)}_u,a^{(\dagger)}_\iota\right]=0
\een

It is not possible to design a root string from $\delta_u-\delta_v<0$ 
to $\al_1$ implying that
\ben
 V_{\delta_u-\delta_v}^{\al_1}=0
\een
This fact is used in deriving (\ref{LJdelta}). However, root strings from 
$-\al^\pm$ to $\al_1$ do exist. They are of the form
\bea
 -\al^-+...+(\eps_2-\delta_u)+\al_1&=&\al_1\nn
 -\al^-+...+(\eps_1-\delta_u)&=&\al_1
\label{rootm}
\eea
where root strings from $-\al^+$ are obtained by ``inserting'' an
additional $\al_1$:
\bea
 -\al^++...+\al_1+...+(\eps_2-\delta_u)+\al_1&=&\al_1\nn
 -\al^++...+(\eps_2-\delta_u)+\al_1+\al_1&=&\al_1\nn
 -\al^++...+\al_1+...+(\eps_1-\delta_u)&=&\al_1\nn
 -\al^++...+(\eps_1-\delta_u)+\al_1&=&\al_1
\label{rootp}
\eea
Possible additions of positive even roots $\delta_{v'}-\delta_{v}$ are 
indicated by ``...''. Let us consider the polynomials (\ref{pol})
\ben
 V_{-\al^\pm}^{\al_1}=\left[\sum_{n\geq0}\frac{1}{n!}(-C)^n
  \right]_{-\al^\pm}^{\al_1}
\een
and compare the two polynomials in order to derive the relation
(\ref{VpVm}). This we do by considering $\frac{1}{n!}(-C)^n$ in
$\Vm$ and $\frac{1}{(n+1)!}(-C)^{n+1}$ in $\Vp$. There are two cases, as 
such terms involve either $c^{\eps_1-\delta_u}$ or $c^{\eps_2-\delta_u}$
for some $u$, and we will discuss them separately. In the first case
the relevant root strings are the lower one in (\ref{rootm}) and the
two lower ones in (\ref{rootp}). Their differences are characterized by the
structure constants
\ben
 {f_{-(\eps_2-\delta_u),(\eps_1-\delta_u)}}^{\al_1}=1
\label{1m}
\een
and
\bea
 {f_{-(\eps_1-\delta_v),\al_1}}^{-(\eps_2-\delta_v)}
 {f_{-(\eps_2-\delta_u),(\eps_1-\delta_u)}}^{\al_1}&=&1\nn
 {f_{-(\eps_1-\delta_u),\eps_1-\delta_u}}^j{f_{j,\al_1}}^{\al_1}&=&1
\label{1p}
\eea 
Now, each time the situation (\ref{1m}) occurs in $\frac{1}{n!}(-C)^n$ in
$\Vm$, the situations (\ref{1p}) occur $n$ times and once, respectively,
in $\frac{1}{(n+1)!}(-C)^{n+1}$ in $\Vp$. This is in accordance with 
(\ref{VpVm}). A similar analysis of the terms involving
$c^{\eps_2-\delta_u}$ leads to the characterizations
\ben
 {f_{-(\eps_2-\delta_u),\eps_2-\delta_u}}^j{f_{j,\al_1}}^{\al_1}=-1
\label{2m}
\een
and
\bea
 {f_{-(\eps_1-\delta_v),\al_1}}^{-(\eps_2-\delta_v)}
  {f_{-(\eps_2-\delta_u),\eps_2-\delta_u}}^j{f_{j,\al_1}}^{\al_1}&=&-1\nn
 {f_{-(\eps_1-\delta_u),\eps_2-\delta_u}}^{-\al_1}
 {f_{-\al_1,\al_1}}^1{f_{1,\al_1}}^{\al_1}&=&-2
\label{2p}
\eea
Each time the situation (\ref{2m}) occurs in $\frac{1}{n!}(-C)^n$ in
$\Vm$, the situations (\ref{2p}) occur $n-1$ times and once, respectively,
in $\frac{1}{(n+1)!}(-C)^{n+1}$ in $\Vp$. However, as the last situation
contributes with a factor $-2$, the terms involving $c^{\eps_2-\delta_u}$
also agree with the relation (\ref{VpVm}), which is thereby proven.

\subsection{Polynomial Relations}

Using that the polynomials $V$ enter in a differential operator
realization of a Lie superalgebra leads to the polynomial relations
\ben
 \fabc V_c^\al=V_a^\nu\pa_\nu V_b^\al-(-1)^{p(a)p(b)}V_b^\nu\pa_\nu
  V_a^\al
\een
as discussed in Ref. \cite{Ras}. The parity $p(a)$ of the index $a$ 
is defined as 1 for $a$ an odd root and 0 otherwise.
Particularly useful are the following relations:
\bea
 V_{\al_1}^\nu\pa_\nu\Vm&=&0\nn
 V_1^\nu\pa_\nu\Vm&=&-\Vm\nn
 V_{-\al_1}^\nu\pa_\nu\Vm&=&-\g\Vm
\label{VVm}
\eea
\bea
 V_{\al^+}^\mu\pa_\mu V_{-\beta^-}^{\al_1}&=&\delta_{\al^-,\beta^-}\nn
 V_{\al^-}^\mu\pa_\mu V_{-\beta^-}^{\al_1}&=&\g\delta_{\al^-,\beta^-}\nn
 V_{-\al^+}^\mu\pa_\mu V_{-\beta^-}^{\al_1}&=&-V_{-\al^-}^{\al_1}
  V_{-\beta^-}^{\al_1}\nn
 V_{-\al^-}^\mu\pa_\mu V_{-\beta^-}^{\al_1}&=&0
\eea
\bea
 {f_{\al^-,-\beta^-}}^cV_c^{\al_1}&=&\g\delta_{\al^-,\beta^-}\nn
 {f_{\al^-,-\beta^-}}^cV_c^{\mu}\pa_\mu V_{-\nu^-}^{\al_1}&=&
  \delta_{\al^-,\nu^-}V_{-\beta^-}^{\al_1}
\label{fVc}
\eea
\bea
 {f_{\al^+,-\beta^-}}^cJ_c&=&\delta_{\al^-,\beta^-}E_1\nn
 {f_{\al^+,-\beta^+}}^cJ_c&=&\delta_{\al^-,\beta^-}H_1+
  {f_{\al^-,-\beta^-}}^cJ_c\nn
 {f_{\al^-,-\beta^+}}^cJ_c&=&\delta_{\al^-,\beta^-}F_1
\eea 
\bea
 {f_{\al^-,-\beta^-}}^c{f_{c,\la^-}}^dJ_d&=&\delta_{\al^-,\beta^-}J_{\la^-}
  -\delta_{\beta^-,\la^-}J_{\al^-}\nn
 {f_{\al^-,-\beta^-}}^c{f_{c,\la^+}}^dJ_d&=&
  -\delta_{\beta^-,\la^-}J_{\al^+}\nn
 {f_{\al^-,-\beta^-}}^c{f_{c,-\la^-}}^dJ_d&=&\delta_{\al^-,\la^-}
  J_{-\beta^-}-\delta_{\al^-,\beta^-}J_{-\la^-}\nn
 {f_{\al^-,-\beta^-}}^c{f_{c,-\la^+}}^dJ_d&=&\delta_{\al^-,\la^-}
  J_{-\beta^+}\nn
 {f_{\al^-,-\beta^-}}^c{f_{c,\al_1}}^dJ_d&=&-\delta_{\al^-,\beta^-}E_1\nn
 {f_{\al^-,-\beta^-}}^c{f_{c,1}}^dJ_d&=&0\nn
 {f_{\al^-,-\beta^-}}^c{f_{c,-\al_1}}^dJ_d&=&\delta_{\al^-,\beta^-}F_1
\eea
\ben
 {f_{\al^-,-\beta^-}}^c{f_{\mu^-,-\nu^-}}^d{f_{c,d}}^eJ_e=
  \delta_{\al^-,\nu^-}{f_{\mu^-,-\beta^-}}^cJ_c
  -\delta_{\mu^-,\beta^-}{f_{\al^-,-\nu^-}}^cJ_c
\een
In deriving some of these relations we have made use of the explicit
oscillator realization (\ref{osc}).

\end{document}